\documentclass[floatfix,twocolumn,showpacs,preprintnumbers,amsmath,amssymb,prc,superscriptaddress]{revtex4} 

\usepackage{graphicx}
\usepackage{dcolumn}
\usepackage{bm}

\newcommand{\bef}{\begin{figure}[!htb]}
\newcommand{\eef}{\end{figure}}

\newcommand{\be}{\begin{equation}}
\newcommand{\ee}{\end{equation}}
\newcommand{\bea}{\begin{eqnarray}}
\newcommand{\eea}{\end{eqnarray}}

\newcommand{\pt}{$p_T$ }

\def \GeVc {\mbox{$\mathrm{GeV}/c$}}

\usepackage{color}

\begin{document}

\title{Identified particle production, azimuthal anisotropy, and interferometry measurements in Au+Au 
collisions at $\sqrt{\bm {s_{NN}}} =$ 9.2 GeV}

\affiliation{Argonne National Laboratory, Argonne, Illinois 60439, USA}
\affiliation{University of Birmingham, Birmingham, United Kingdom}
\affiliation{Brookhaven National Laboratory, Upton, New York 11973, USA}
\affiliation{University of California, Berkeley, California 94720, USA}
\affiliation{University of California, Davis, California 95616, USA}
\affiliation{University of California, Los Angeles, California 90095, USA}
\affiliation{Universidade Estadual de Campinas, Sao Paulo, Brazil}
\affiliation{University of Illinois at Chicago, Chicago, Illinois 60607, USA}
\affiliation{Creighton University, Omaha, Nebraska 68178, USA}
\affiliation{Czech Technical University in Prague, FNSPE, Prague, 115 19, Czech Republic}
\affiliation{Nuclear Physics Institute AS CR, 250 68 \v{R}e\v{z}/Prague, Czech Republic}
\affiliation{University of Frankfurt, Frankfurt, Germany}
\affiliation{Institute of Physics, Bhubaneswar 751005, India}
\affiliation{Indian Institute of Technology, Mumbai, India}
\affiliation{Indiana University, Bloomington, Indiana 47408, USA}
\affiliation{University of Jammu, Jammu 180001, India}
\affiliation{Joint Institute for Nuclear Research, Dubna, 141 980, Russia}
\affiliation{Kent State University, Kent, Ohio 44242, USA}
\affiliation{University of Kentucky, Lexington, Kentucky, 40506-0055, USA}
\affiliation{Institute of Modern Physics, Lanzhou, China}
\affiliation{Lawrence Berkeley National Laboratory, Berkeley, California 94720, USA}
\affiliation{Massachusetts Institute of Technology, Cambridge, MA 02139-4307, USA}
\affiliation{Max-Planck-Institut f\"ur Physik, Munich, Germany}
\affiliation{Michigan State University, East Lansing, Michigan 48824, USA}
\affiliation{Moscow Engineering Physics Institute, Moscow Russia}
\affiliation{City College of New York, New York City, New York 10031, USA}
\affiliation{NIKHEF and Utrecht University, Amsterdam, The Netherlands}
\affiliation{Ohio State University, Columbus, Ohio 43210, USA}
\affiliation{Old Dominion University, Norfolk, VA, 23529, USA}
\affiliation{Panjab University, Chandigarh 160014, India}
\affiliation{Pennsylvania State University, University Park, Pennsylvania 16802, USA}
\affiliation{Institute of High Energy Physics, Protvino, Russia}
\affiliation{Purdue University, West Lafayette, Indiana 47907, USA}
\affiliation{Pusan National University, Pusan, Republic of Korea}
\affiliation{University of Rajasthan, Jaipur 302004, India}
\affiliation{Rice University, Houston, Texas 77251, USA}
\affiliation{Universidade de Sao Paulo, Sao Paulo, Brazil}
\affiliation{University of Science \& Technology of China, Hefei 230026, China}
\affiliation{Shandong University, Jinan, Shandong 250100, China}
\affiliation{Shanghai Institute of Applied Physics, Shanghai 201800, China}
\affiliation{SUBATECH, Nantes, France}
\affiliation{Texas A\&M University, College Station, Texas 77843, USA}
\affiliation{University of Texas, Austin, Texas 78712, USA}
\affiliation{Tsinghua University, Beijing 100084, China}
\affiliation{United States Naval Academy, Annapolis, MD 21402, USA}
\affiliation{Valparaiso University, Valparaiso, Indiana 46383, USA}
\affiliation{Variable Energy Cyclotron Centre, Kolkata 700064, India}
\affiliation{Warsaw University of Technology, Warsaw, Poland}
\affiliation{University of Washington, Seattle, Washington 98195, USA}
\affiliation{Wayne State University, Detroit, Michigan 48201, USA}
\affiliation{Institute of Particle Physics, CCNU (HZNU), Wuhan 430079, China}
\affiliation{Yale University, New Haven, Connecticut 06520, USA}
\affiliation{University of Zagreb, Zagreb, HR-10002, Croatia}

\author{B.~I.~Abelev}\affiliation{University of Illinois at Chicago, Chicago, Illinois 60607, USA}
\author{M.~M.~Aggarwal}\affiliation{Panjab University, Chandigarh 160014, India}
\author{Z.~Ahammed}\affiliation{Variable Energy Cyclotron Centre, Kolkata 700064, India}
\author{A.~V.~Alakhverdyants}\affiliation{Joint Institute for Nuclear Research, Dubna, 141 980, Russia}
\author{B.~D.~Anderson}\affiliation{Kent State University, Kent, Ohio 44242, USA}
\author{D.~Arkhipkin}\affiliation{Brookhaven National Laboratory, Upton, New York 11973, USA}
\author{G.~S.~Averichev}\affiliation{Joint Institute for Nuclear Research, Dubna, 141 980, Russia}
\author{J.~Balewski}\affiliation{Massachusetts Institute of Technology, Cambridge, MA 02139-4307, USA}
\author{O.~Barannikova}\affiliation{University of Illinois at Chicago, Chicago, Illinois 60607, USA}
\author{L.~S.~Barnby}\affiliation{University of Birmingham, Birmingham, United Kingdom}
% ...

\collaboration{STAR Collaboration}\noaffiliation

\date{\today}
\begin{abstract}
We present the first measurements of identified hadron production, 
azimuthal anisotropy, and pion interferometry from Au+Au collisions 
below the nominal injection energy at the Relativistic Heavy-Ion Collider 
(RHIC) facility. The data were collected using the large acceptance STAR 
detector at $\sqrt{s_{NN}}=$ 9.2 GeV from a test run of the collider 
in the year 2008. Midrapidity results on 
multiplicity density ($dN/dy$) in rapidity ($y$),
average transverse 
momentum ($\langle p_{T} \rangle$), particle ratios, elliptic flow, 
and HBT radii are consistent with the corresponding results at 
similar $\sqrt{s_{NN}}$ from fixed target experiments. Directed flow 
measurements are presented for both midrapidity and forward rapidity regions.
Furthermore the collision centrality dependence of identified 
particle $dN/dy$, $\langle p_{T} \rangle$, and particle ratios are discussed.
These results also demonstrate that the capabilities of the STAR
detector, although optimized for $\sqrt{s_{NN}}=$ 200 GeV, are suitable
for the proposed QCD critical point search and exploration of the QCD
phase diagram at RHIC.

\end{abstract}
\maketitle

\section{Introduction}

Exploring the Quantum Chromodynamics (QCD) phase diagram is one of the goals of 
high energy heavy-ion collision experiments~\cite{starwhitepaper}.       
The QCD phase diagram is usually plotted as temperature ($T$) versus baryon chemical 
potential ($\mu_{\mathrm {B}}$). Assuming a thermalized system is reached in heavy-ion 
collisions, both of these quantities can be varied by 
changing the collision energy~\cite{thermalmodels}. The phase diagram shows
a possible transition from a high energy density and 
high temperature phase dominated by partonic degrees of freedom, 
to a phase where the relevant degrees of freedom are hadronic~\cite{phasedia}. 
Several observations at the top 
RHIC energy, such as the suppression of high transverse momentum 
($p_{T}$) hadron production in Au+Au collisions relative to
$p$+$p$ collisions~\cite{starraa}, large elliptic flow ($v_{2}$) for hadrons 
with light, as well as heavier strange valence quarks, and differences between 
baryon and meson $v_{2}$ at intermediate $p_{T}$ for Au+Au collisions, 
have been associated with the existence of a phase with partonic degrees of freedom 
in the initial stages of heavy-ion collisions~\cite{starwhitepaper, starraa, starv2}. 
Lowering the collision energy and studying the energy dependence of these observables 
will allow us to search as a function of center of mass energy ($\sqrt{s_{NN}}$) or 
($T$, $\mu_{\mathrm {B}}$) for 
the onset of the transition to 
a phase with partonic degrees of freedom at the early stage of the collision.

Lattice QCD calculations indicate that 
the system produced at $\mu_{\mathrm {B}} =$ 0 evolves through a rapid
crossover in the quark-hadron phase transition~\cite{crossover}. Calculations 
from lattice QCD~\cite{firstorder} and those from several QCD-based models~\cite{qcdmodel} 
suggest that for collisions corresponding 
to large $\mu_{\mathrm {B}}$,
the transition is first order. The point in the ($T$, $\mu_{\mathrm {B}}$)
plane where the first order phase transition ends, is the QCD critical point~\cite{qcdcp}. 
Theoretical predictions of the location of this point on the phase diagram are subject to  
various ambiguities~\cite{bmqm09}.
An experimental program for locating the QCD critical point through its signatures~\cite{bmqm09,cpsig} 
(e.g., long range fluctuations in event-by-event observables) is one of the 
exciting possibilities at the
RHIC facility. These motivations form the basis of the proposal~\cite{bes} by the experiments 
at RHIC to carry out a detailed program  of exploring the phase 
diagram by varying the collision energy in high energy heavy-ion collisions. 

As an initial step to test the capabilities of the collider and experiments, a short
run was conducted in the year 2008 at RHIC. 
The Au ions were collided at $\sqrt{s_{NN}} =$ 9.2 GeV, which is below the
injection energy of $\sqrt{s_{NN}} =$ 19.6 GeV. At and below nominal injection energy, 
RHIC runs as a colliding storage ring, further 
details of which can be found in Ref.~\cite{satogata}.
The data taking period
lasted for less than five hours at the Solenoidal Tracker
at RHIC (STAR) experiment. This paper presents results based 
on the analysis of this small data set and demonstrates the success of 
the test run in achieving its objectives. The measurements shown here 
are the first step towards a detailed exploration of the QCD phase diagram at 
RHIC.

The paper is organized as follows: 
The next section briefly presents the detectors used and details of the data analysis. 
In section III, we present the results including $p_{T}$ spectra,  
$dN/dy$,
$\langle p_{T} \rangle$ and particle ratios as a function of 
collision centrality and  $\sqrt{s_{NN}}$. We also discuss results 
on directed flow ($v_{1}$), elliptic flow ($v_{2}$), and pion interferometry in this 
section. In section IV, we discuss the freeze-out conditions.
Finally, in section V we summarize the results and provide a brief outlook
of the upcoming Beam Energy Scan program at RHIC.

\section{Experiment and Data Analysis}

\subsection{STAR detector}

The results presented here are based on data taken 
at STAR~\cite{starnim} 
in Au+Au collisions at $\sqrt{s_{NN}}$ = 9.2 GeV. 
This data set 
is taken with a minimum bias trigger. 
The trigger detectors used are the Beam-Beam 
Counter (BBC) and Vertex Position Detector (VPD)~\cite{vpd}. 
The BBCs are scintillator 
annuli mounted around the beam pipe beyond the east and west 
pole-tips of the STAR magnet at about 375 cm from the 
center of the nominal interaction region (IR). 
The inner tiles of the BBCs, with a pseudorapidity ($\eta$) range of 3.8 $< |\eta| <$ 5.2 and
full azimuthal coverage $\Delta\phi = 2\pi$, 
are used to reconstruct the 
first-order event plane for the directed flow analysis.
The VPDs are based on the conventional technology of plastic scintillator read-out by 
photomultiplier tubes. They consist of  two identical detector assemblies
very close to the beam pipe, one on each side at a distance of
$|V_z|=$ 5.6 m from the center of the IR.
The main detector used to obtain
the results on $p_{\mathrm {T}}$ spectra, yields, particle ratios, azimuthal anisotropy 
parameters, 
and pion interferometry for charged hadrons is the Time Projection Chamber 
(TPC)~\cite{startpc}. 
The TPC is the primary tracking device at STAR. It is 4.2 m long 
and 4 m in diameter. 
Its acceptance covers $\pm1.8$ units of pseudorapidity ($\eta$) 
and the full azimuthal angle. 
The sensitive volume of the TPC contains P10 gas (10\% methane, 
90\% argon) regulated at 2 mbar  above atmospheric pressure. The TPC data are used to 
determine particle trajectories, momenta, and particle-type through ionization energy 
loss ($dE/dx$). 
STAR's solenoidal magnet field used for this low energy Au+Au test run was 0.5\,T.
In addition we present directed flow measurements from forward rapidities.
These results used the data taken by the Forward Time Projection 
Chambers (FTPCs)~\cite{starftpc}. There are two FTPCs located around the beam axis 
on both sides of the collision point. The sensitive medium is a gas mixture of 
equal parts Ar and CO$_2$ by weight. The FTPCs detect charged particles 
in the pseudorapidity region 2.5 $\le |\eta| \le$ 4.0, with full azimuthal coverage.
The details of the design and other characteristics of the STAR detectors 
can be found in Ref.~\cite{starnim}.

\bef
\includegraphics[scale=0.39]{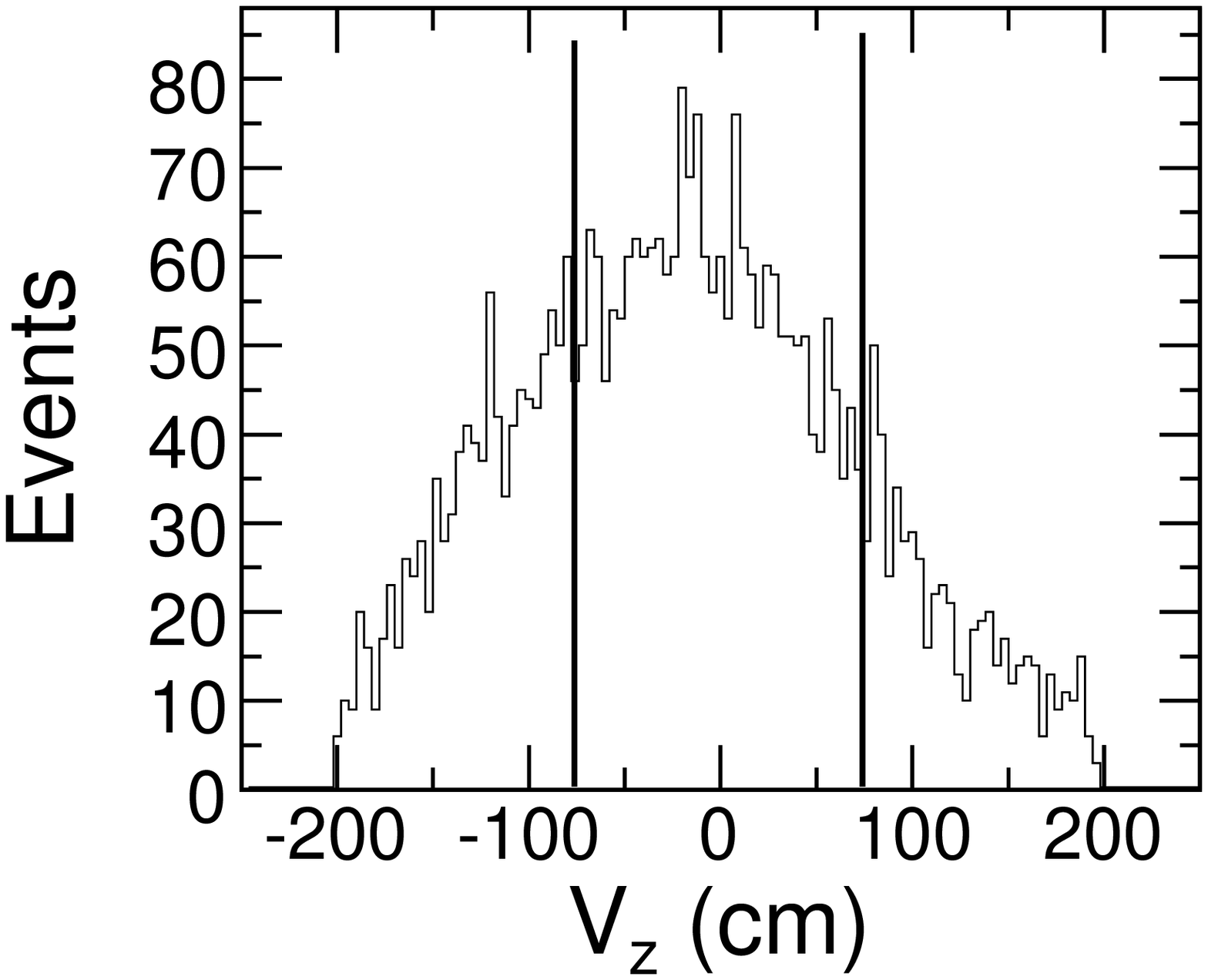}
\caption{Event-by-event distribution of
the $z$-position of the primary vertex ($V_{z}$) in Au+Au collisions 
at $\sqrt{s_{NN}}$ = 9.2 GeV. The vertical solid lines show the condition
of $|V_{z}|$ $<$ 75 cm for selected events.}
\label{vz}
\eef
\subsection{Event selection}
The primary vertex for each minimum bias event is determined by finding the
best point of common origin of the tracks measured in the TPC. The distribution of the 
primary vertex position along the longitudinal beam direction ($V_{z}$) is 
shown in Fig.~\ref{vz}. The distribution is a broad 
Gaussian varying between $-$200 and 200 cm, with a root mean square deviation of 89 cm. 
Only those events which have a $V_{z}$ within 75 cm of the nominal collision 
point (center of the detector) are selected for the analysis, corresponding 
to 57\% of the total events 
recorded. 
This value is chosen by the trade-off between uniform detector performance 
within $|\eta|$ $<$ 1.0 
and sufficient statistical significance of the measured observables.
In order to reject events which involve interactions with the beam pipe and beam-gas 
interactions, the event vertex radius
(defined as $\sqrt{V_{x}^{2} + V_{y}^{2}}$ where 
$V_{x}$ and $V_{y}$ are the vertex positions along the $x$ and $y$ directions) 
is required to be less than 2 cm. The $V_{x}$ vs. $V_{y}$ distribution is shown in
Fig.~\ref{vrad}. The circle with dotted lines corresponds to the event vertex
radius of 2 cm.
A total of about 3000 events pass the selection criteria described above.
\bef
\includegraphics[scale=0.39]{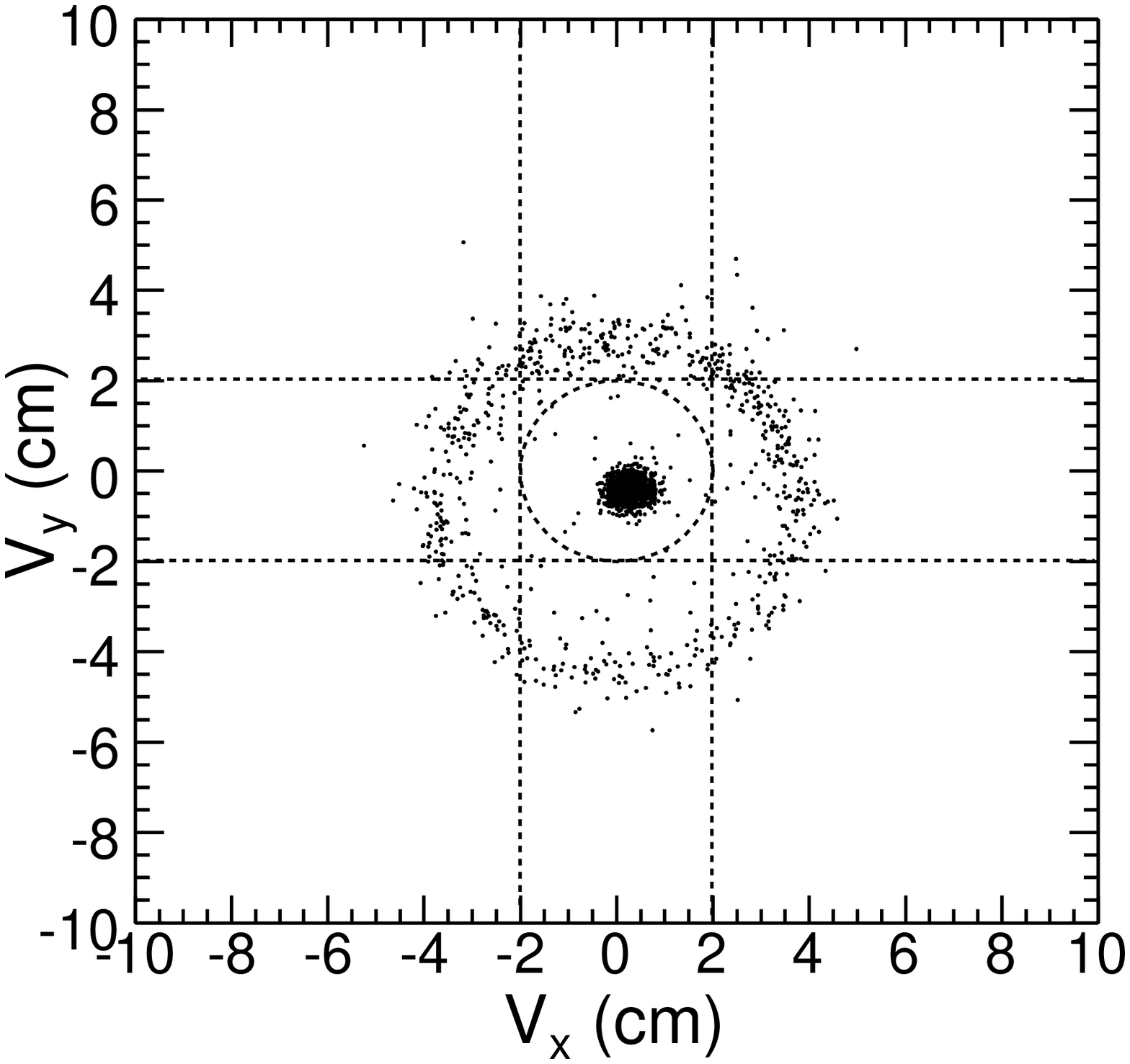}
\caption{Event-by-event distribution of
$V_{x}$ vs. $V_{y}$ in Au+Au collisions 
at $\sqrt{s_{NN}}$ = 9.2 GeV. The circle with dotted lines 
corresponds to a radius (= $\sqrt{V_{x}^{2} + V_{y}^{2}}$) of 
2 cm.}
\label{vrad}
\eef

\subsection{Centrality selection}
Centrality classes in Au+Au collisions at $\sqrt{s_{NN}}$ = 9.2 GeV are defined 
using the number of charged particle tracks reconstructed in the main TPC 
over the full azimuth,
pseudorapidity $|\eta| < 0.5$ and $|V_{z}| < $ 75 cm.

Figure~\ref{centrality} shows the uncorrected multiplicity distribution
for charged tracks from the real data 
($N_{\mathrm {ch}}^{\mathrm {TPC}}$, open circles) and for those 
obtained from simulation (dashed histogram).
Simulated multiplicity density is calculated using the two-component model~\cite{kharznard} 
with the number of 
participants ($N_{\rm part}$) and number of collisions ($N_{\rm coll}$) extracted from 
the Glauber Monte Carlo simulation as
\begin{equation}
\frac{dN_{\rm{ch}}}{d\eta} = n_{pp}\left[ (1-x) \frac{N_{\rm part}}{2} + xN_{\rm coll} \right].
\end{equation}
Here $n_{pp}$ is the average multiplicity in minimum bias $p$+$p$ collisions and $x$ is the 
fraction of the 
hard component.
The inelastic cross-section for $p$+$p$ collisions used in the Glauber Model simulations is 
31.5 mb~\cite{pdg}. 
The event-by-event multiplicity fluctuation has been taken into
account by convoluting the Negative Binomial Distributions (NBD) 
for a given $N_{\rm part}$ and $N_{\rm coll}$. 
The NBD distribution in multiplicity $n$ has two parameters, $n_{pp}$ and $k$, and is defined as,
\begin{equation}
P_{\rm NBD}(n_{pp}, k; n) 
= \frac{\Gamma(n+k)}{\Gamma(n+1)\Gamma(k)} \cdot \frac{(n_{pp}/k)^n}{(n_{pp}/k + 1)^{n+k}},
\end{equation}
where $\Gamma$ is the Gamma function. The values $k = 2.1$ and $n_{pp}= 1.12$
are obtained by fitting
the measured multiplicities with those 
from the simulation.
The simulated multiplicity distribution is not sensitive to the $k$ 
parameter. The distributions are found to be similar for varying $k$ values such 
as $k=$ 1.0, 1.6, and 3.0.
The fitting is performed for $N_{\rm {ch}} > 17$ in order 
to avoid the trigger inefficiency in peripheral collisions. The $x$ value is fixed 
at 0.11 $\pm$ 0.03, obtained by extrapolating data from the PHOBOS collaboration~\cite{phobos}.
The centrality is defined by calculating the fraction of the total cross-section 
obtained from the simulated multiplicity. 
\bef
\includegraphics[scale=0.39]{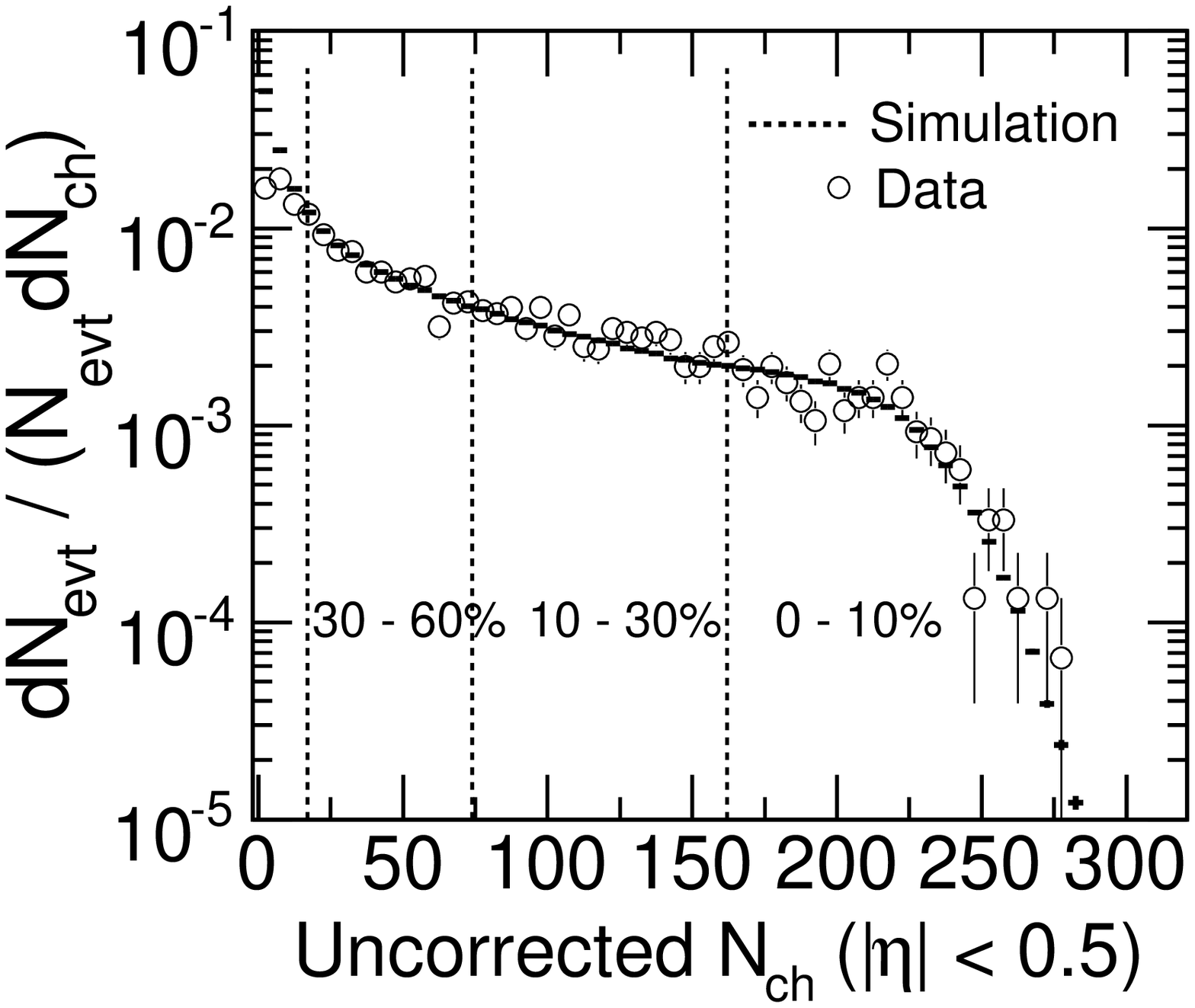}
\caption{ 
Uncorrected charged particle multiplicity distribution (open circles) 
measured in the TPC within $|\eta| < 0.5$ in Au + Au collisions at $\sqrt{s_{NN}} =$ 9.2 GeV.
The dashed histogram represents the simulated multiplicity distribution. 
The vertical dashed lines reflect the centrality selection criteria used in the paper.
Errors are statistical only.}
\label{centrality}
\eef

Table~\ref{table1} lists the 
centrality selection criteria for Au+Au collisions at $\sqrt{s_{NN}} =$ 9.2 GeV.
We have divided the events
into three centrality classes, 0--10\%, 10--30\%, and 30--60\% of the total cross-section. The 
mean values of 
$N_{\rm part}$ and $N_{\rm coll}$ have been evaluated for these centrality bins and are given
in Table~\ref{table1}. Systematic uncertainties on $\langle N_{\rm part} \rangle$ and 
$\langle N_{\rm coll} \rangle$ 
have been estimated by varying $n_{pp}$ and $x$ in the two-component 
model as well as varying the input parameters in the Glauber Monte Carlo simulation. 
The final errors on $\langle N_{\rm part} \rangle$ and 
$\langle N_{\rm coll} \rangle$ are the quadrature sum of these individual systematic errors. 
The results presented in this paper cover the collision centrality range of 0--60\%. 
The results from more peripheral collisions are not presented due to
large trigger inefficiencies in this test run.

\begin{table}
\caption{ Centrality selection, average number of participating nucleons 
($\langle N_{\mathrm{part}} \rangle$), and
average number of binary collisions ($\langle N_{\mathrm{coll}} \rangle$).
\label{table1}}
\vspace{0.15cm}
\begin{tabular}{c|c|c|c}
\hline
\% cross-section&$N_{\mathrm{ch}}^{\mathrm{TPC}}$ &$\langle N_{\mathrm{part}} \rangle$
&$\langle N_{\mathrm{coll}} \rangle$\\
\hline
0--10  &  $>$ 162   &~~ 317 $\pm$ 4   &~~ 716 $\pm$ 83\\
10--30 &  74--162   &~~ 202 $\pm$ 11  &~~ 395 $\pm$ 34\\
30--60 &  17--74    &~~ 88  $\pm$ 10  &~~ 133 $\pm$ 20\\
\hline
\end{tabular}
\end{table}

\subsection{Track selection and particle identification}
Track selection criteria for the various analyses are presented in Table~\ref{table2}.
In order to avoid admixture of tracks from secondary vertices, a requirement is placed on
the distance of closest approach (DCA) between each track and the event vertex. In order to
prevent multiple counting of split tracks, a condition is placed on the number of track points
($N_{\rm{fit}}$) 
used in the reconstruction of the track. 
Tracks can have a maximum of 45 hits in the TPC.

\begin{table}
\caption{Track selection criteria for various analyses presented in this paper.
\label{table2}}
\vspace{0.15cm}
\begin{tabular}{c|c|c|c|c}
\hline
Analysis         & DCA (cm)      & $N_{\rm{fit}}$ & $\eta$ or $y$ & $p_{\mathrm {T}}$ (GeV/$c$)\\
\hline
$p_{\mathrm {T}}$ spectra  &  $<$ 3   & $>$ 20      & $|y|$    $<$ 0.5         & $>$ 0.1   \\
$v_{1}$(TPC)     &  $<$ 1   & $>$ 20      & $|\eta|$ $<$ 1.3         & 0.15 -- 2.0 \\
$v_{1}$(FTPC)    &  $<$ 1   & $>$ 5       & 2.5 $<$ $|\eta|$ $<$ 4.0 & 0.15 -- 2.0   \\
$v_{2}$          &  $<$ 3   & $>$ 15      & $|\eta|$ $<$ 1.0         & 0.1 -- 2.0 \\
HBT              &  $<$ 3   & $>$ 15      & $|y|$ $<$ 0.5             & $k_{\rm {T}}$ : \\
                 &            &             &                           & 0.15--0.25 \\
\hline
\end{tabular}
\end{table}

\bef
\includegraphics[scale=0.39]{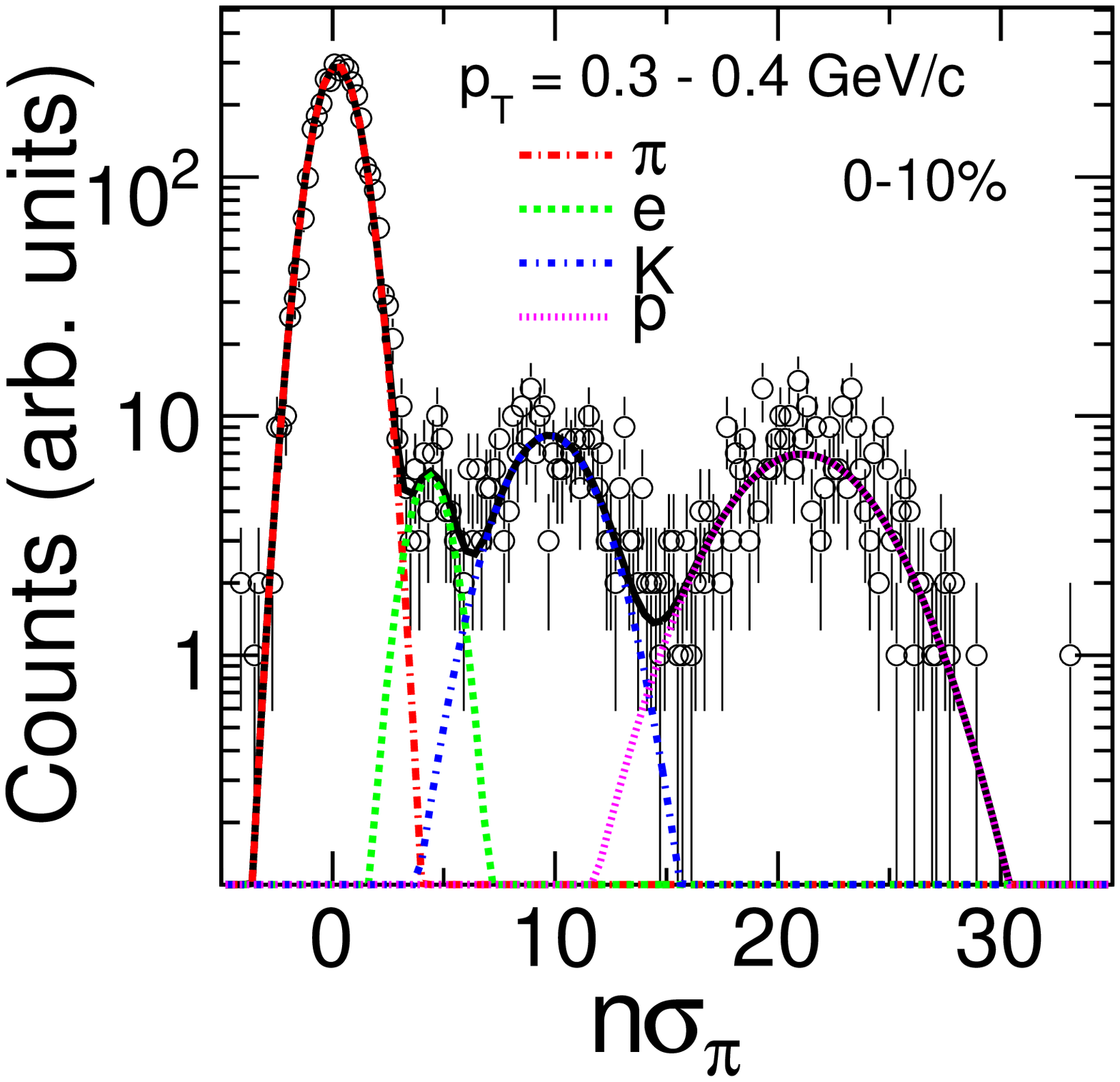}
\caption{ (Color online) $dE/dx$ distribution for positively charged hadrons in the TPC,
normalized by the expected pion $dE/dx$ at 0.3~$<$~$p_{T}$~$<$~0.4~GeV/$c$ and
$\mid$$y$$\mid$~$<$~0.5 in Au+Au collisions at $\sqrt{s_{NN}}$ = 9.2 GeV. The 
curves are Gaussian fits representing contributions from pions (dot-dashed, red), electrons 
(dashed, green), kaons (dot-dashed, blue), and protons (dotted, magenta). See text for details.
Errors are statistical only.
}
\label{nsigma}
\eef
To extract the pion yield in a given $p_{T}$ bin, we
perform an eight-Gaussian fit to the normalized $dE/dx$ distributions of
positively charged and negatively charged hadrons, simultaneously.
The normalized $dE/dx$ in general is defined as
\begin{equation}
\label{eqnsigma}
n\sigma_{X}=\frac{\log((dE/dx)/B_{X})}{\sigma_{X}}, 
\end{equation}
where $X$
is the particle type ($e^{\pm},\pi^{\pm},K^{\pm}$, $p$, or $\bar{p}$), $B_{X}$ is the
expected mean $dE/dx$ of  particle $X$, and $\sigma_{X}$ is the
$dE/dx$ resolution of the TPC, which is a function of the track length in the TPC.
The expected mean $dE/dx$ of  particle $X$
is calculated using a Bichsel function for the energy loss in thin
layers of P10 in the STAR TPC~\cite{startpc,bic}. Good agreement between the measurement
and the calculation 
was demonstrated previously~\cite{rdEdx}.
Figure~\ref{nsigma} 
shows a typical $dE/dx$ distribution normalized to the pion $dE/dx$ 
(referred to as the $n\sigma_{\pi}$ distribution) for charged hadrons 
with 0.3~$<$~$p_{T}$~$<$~0.4~GeV/$c$ and $\mid$$y$$\mid$ $<$ 0.5.
The counts under the Gaussian about $n\sigma_{\pi}$ $\sim$ 0 give the yield
of pions for a particular $p_{T}$ range. A similar procedure is followed to
obtain yields for other $p_{T}$ ranges and for yields of kaons and protons. 
Further details of extracting raw yields of identified hadrons from normalized $dE/dx$
distributions 
can be found in Ref.~\cite{STARPID}.

For the elliptic flow analysis of identified hadrons, the 
criteria of $|n\sigma_{\pi}|$ $<$ 2 and $|n\sigma_{p}|$ $<$ 2 are used for extracting 
pion and proton 
$v_{2}$. Since the measurements are carried out at low $p_{T}$ ($<$ 1.0 GeV/$c$), such an identification 
criterion is reasonable. For the pion interferometry analysis, the particle identification 
conditions are $|n\sigma_{\pi}|$ $<$ 2, $|n\sigma_{p}|$ $>$ 2, and  $|n\sigma_{K}|$ $>$ 2, and
the average transverse momentum ($k_{T}$ $=$ ($|\vec{p}_{\rm 1T}$ $+$ $\vec{p}_{\rm 2T}|$)$/$2)
is required to fall in the range 150--250 MeV/$c$.

\subsection{Event plane for azimuthal anisotropy}

Azimuthal anisotropy can be quantified by studying the Fourier expansion of the 
azimuthal angle ($\phi$) distribution of produced particles with respect 
to the reaction plane angle ($\Psi_{R}$)~\cite{Methods}. 
The various (order $n$) coefficients in this expansion are defined as:
\begin{equation}
v_{n}=\langle\cos[n(\phi-\Psi_{R})]\rangle.
\end{equation}
The angular brackets in the definition denote an average over 
many particles and events.
Directed flow can be quantified by the first coefficient ($v_{1}$) and
elliptic flow by the second coefficient ($v_{2}$), obtained using the
above equation.

In the azimuthal anisotropy analysis, $v_{1}$ and $v_{2}$ are obtained from the
following procedure. The event flow vector ($Q_n$) and the event plane
angle ($\Psi_n$) are defined by~\cite{Methods}
\begin{equation} \label{Qnx} Q_n\cos(n\Psi_n)\ =\ Q_{nx}\ =\ 
\sum_{i}w_i\cos(n\phi_i),
\end{equation}
\begin{equation} \label{Qny} Q_n\sin(n\Psi_n)\ =\ Q_{ny}\ = \
\sum_{i}w_i\sin(n\phi_i), 
\end{equation}
\begin{equation} \label{Psi} \Psi_n\ =\
\left(\tan^{-1}\frac{Q_{ny}}{Q_{nx}}\right)/n,
\end{equation}
where sums go over all particles $i$ used in the event
plane calculation, and $\phi_i$ and $w_i$ are the laboratory azimuthal angle
and the weight for the $i$-th particle, respectively. 
Tracks used for the calculation of $v_{n}$ are excluded from the calculation of 
the event plane to remove self-correlation effects.

Since finite multiplicity limits the angular resolution of the
reaction plane reconstruction, 
the $v_n^{\rm{obs}}$ has to be corrected for the event
plane resolution by 
\begin{equation} \label{v2EP2} v_{n}\ =\
\frac{v_n^{\rm{obs}}}{\langle \cos[n(\Psi_n-\Psi_R)]\rangle}
\end{equation}
to obtain the real $v_n$, where angular brackets denote an average over a large event sample.
The event plane
resolution is estimated from the correlation of the event planes of
two sub-events. 
Assuming the pure flow correlations between the sub-events, 
the event plane resolution is given by
\begin{equation}
\langle \cos[n(\Psi_{n}^{A}-\Psi_{R})] \rangle = \sqrt { \langle
\cos[n(\Psi_{n}^{A}-\Psi_{n}^{B})] \rangle }, 
\label{subEPres}
\end{equation}
where A and B denote two subgroups of tracks. In this analysis, we
use two random sub-events with equal numbers of particles. 
The full event plane resolution is obtained from the resolution of
the sub-events by
\begin{equation} \langle
\cos[n(\Psi_{n}-\Psi_{R})] \rangle = C \langle
\cos[n(\Psi_{n}^{A}-\Psi_{R})] \rangle, 
\label{EPres}
\end{equation}
where $C$ is a constant calculated from the known 
dependence of the resolution on multiplicity~\cite{Methods}.

For the elliptic flow measurements presented in this paper, 
the TPC tracks are used to reconstruct the reaction plane~\cite{Methods}. 
The weights are taken to be the value of \pt in \GeVc\: up to 
2 \GeVc\, and then constant at 2.0 for \pt $>$ 2 GeV/$c$.
Such weight values are chosen as $v_{2}$ increases with $p_{\mathrm T}$ 
up to 2 GeV/$c$ and then tends to saturates beyond $p_{\mathrm T}$ = 2 GeV/$c$.
The variation of event plane resolution with collision centrality
is shown in Fig.~\ref{evplane}. The values
of the resolution depend on the multiplicity and flow observed in
the events. The resolution values are lower for $\sqrt{s_{NN}}$ = 9.2 GeV,
compared to collisions at $\sqrt{s_{NN}}$ = 200 GeV for similarly defined collision
centrality classes~\cite{flow1}. 
A similar procedure for correcting the observed flow values with the
resolution factor is followed for $v_{1}$ measurements.
The $v_{1}$ results presented here are obtained using two different 
methods: the mixed harmonics
and the standard methods.

In the mixed harmonics method, $v_{1}$ is calculated using 
mixed harmonics involving the second-harmonic event plane~\cite{flow1}.
This method utilizes the large elliptic flow signal, and at the same time 
suppresses the non-flow contributions arising from the correlation of particles from the 
same harmonics. 
The method uses the second order event plane from the TPC ($\Psi_{2}^{\rm {TPC}}$) and the
first order event plane from random sub-events in the 
FTPCs ($\Psi_{1}^{\rm {FTPC1}}$ and $\Psi_{1}^{\rm {FTPC2}}$).
The average resolution for the event plane (as defined in Eqs.~(\ref{subEPres}) and (\ref{EPres})) 
reconstructed from the TPC is 0.46 $\pm$ 0.03, while that 
reconstructed from the FTPCs is 0.41 $\pm$ 0.03, for 0--60\% collision centrality.
The mixed harmonics method is denoted by $v_{1}\{ \rm {EP_{1}}, \rm {EP_{2}}\} $~\cite{flow1}, 
as given below:
\begin{equation}
v_{1}\{ \rm{EP_{1}},\rm{EP_{2}} \} = ~~~~~~~~~~~~~~~~~~~~~~~~~~~~~~~~~~~~~~~~~~~~~~~~~~~\
\nonumber
\end{equation}
\begin{equation}
 \frac{ \langle \cos(\phi+\Psi_{1}^{\rm{FTPC}}-2\Psi_{2}^{\rm{TPC}})\rangle}
{\sqrt{\langle\cos(\Psi_{1}^{\rm{FTPC1}}+\Psi_{1}^{\rm{FTPC2}}-2\Psi_{2}^{\rm{TPC}})\rangle
\rm{Res}(\Psi_{2}^{\rm{TPC}})}},
\end{equation}

\noindent where the emission angle of the particle ($\phi$) is correlated with the
$\Psi_{1}^{\rm{FTPC}}$ of the random sub-event composed of tracks from both the FTPCs
excluding that particle.

In the standard method, the first-order event plane is reconstructed separately
from the FTPC tracks ($v_{1}\{ \rm{EP_{1}}, \rm{FTPC}\}$) and from the BBC hits 
($v_{1}\{ \rm{EP_{1}}, \rm{BBC}\}$).
The event plane reconstructed from 
the detector on one side of the collision (east or west) is called a sub-event plane.
A combination of the east and west sub-event plane vectors provides the full event plane.
In the $v_{1}\{ \rm{EP_{1}}, \rm{FTPC}\}$ method, we used the event plane obtained from the full
FTPC region to obtain the directed flow values measured in the TPC range ($|\eta| <$ 1.3).
A self-correlation arises if $v_{1}$ is obtained using particles from the same 
pseudorapidity region as is used for the event plane reconstruction. 
To avoid this self-correlation in the $v_{1}\{ \rm{EP_{1}}, \rm{FTPC}\}$ method, $v_{1}$ is 
obtained in the east FTPC ($-4.2 < \eta < -2.5$) by using the event plane reconstructed 
in the west FTPC (2.5 $< \eta <$ 4.2), and 
{\it vice versa}.
In the $v_{1}\{ \rm{EP_{1}}, \rm{BBC}\}$ method, the event plane is obtained from the full BBC
region (3.8 $< |\eta| <$ 5.2). Particles used for the estimation of $v_{1}$ with
respect to the BBC full event plane 
cover the $\eta$ range up to 3.8, in order to avoid the self-correlation.
The average resolution of the first order event plane 
for $v_{1}\{ \rm{EP_{1}}, \rm{FTPC}\}$ is 0.41 $\pm$ 0.03 
for 0--60\% central collisions while 
that for $v_{1}\{ \rm{EP_{1}}, \rm{BBC}\}$ is 0.24 $\pm$ 0.07.

\bef
\includegraphics[scale=0.39]{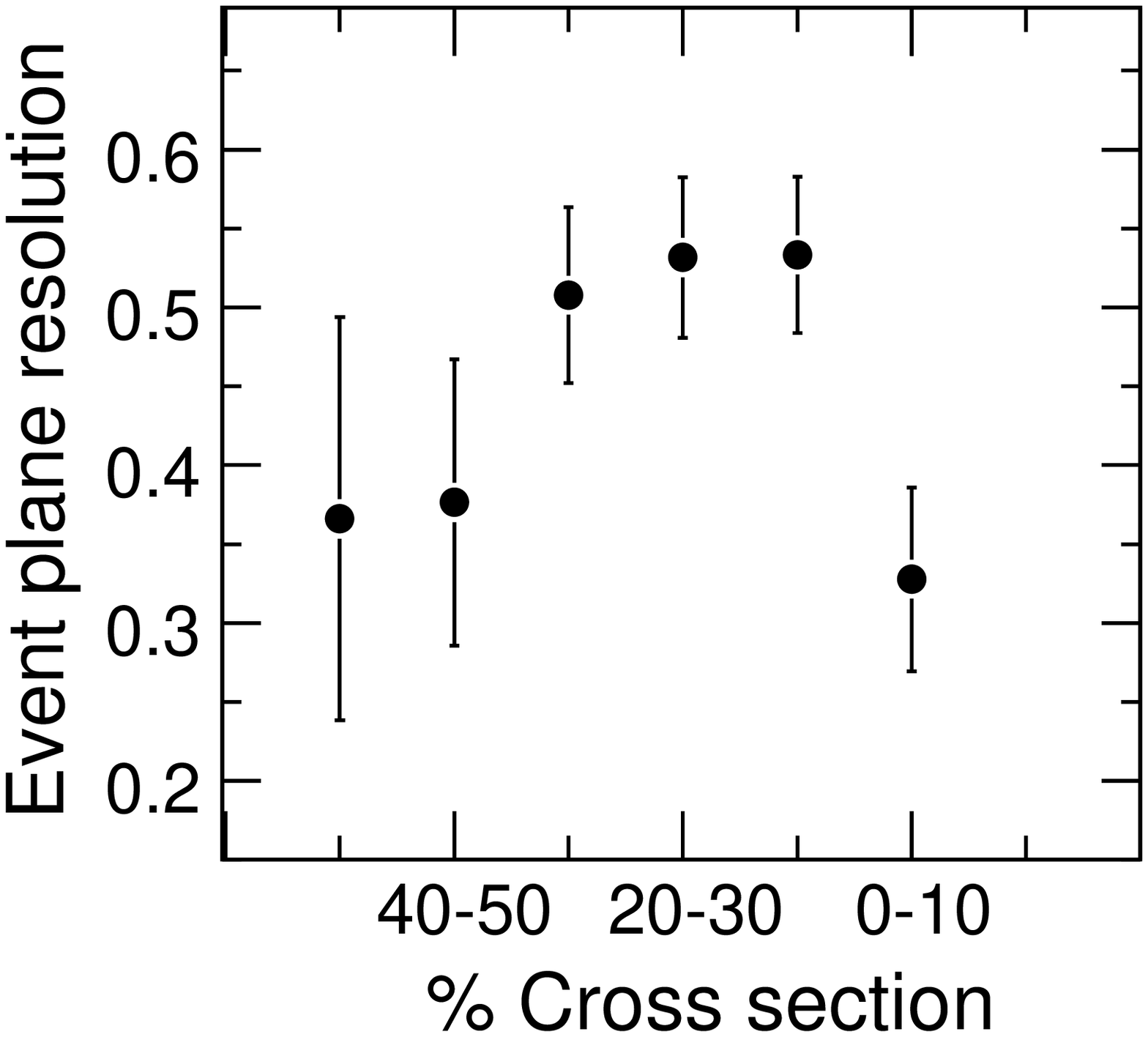}
\caption{ Second order event plane resolution measured in the TPC as a function of
collision centrality for Au+Au collisions at $\sqrt{s_{NN}}$ = 9.2 GeV. 
Errors are statistical only.}
\label{evplane}
\eef

\subsection{Correlation function in $\bm{\pi}$ interferometry}
Experimentally, the two-particle correlation function is obtained from the ratio,
\begin{equation}
\label{eq:one}
C( \vec{q},\vec{k}) = 
\frac{A( \vec{q},\vec{k})}{B(\vec{q},\vec{k})}~~,
\end{equation}
where $A$($\vec{q},\vec{k}$) 
is the distribution of particle pairs with 
relative momentum $\vec{q} = \vec{p_1} - \vec{p_2}$ 
and average momentum $\vec{k} = (\vec{p_1} + \vec{p_2})/2$ 
from the same event, and $B$($\vec{q},\vec{k}$) 
is the corresponding distribution for pairs of particles taken from 
different events~\cite{Kopylov:1972qw,Heinz:1999rw}. The correlation function
is normalized to unity at large $\vec{q}$. 
In the mixed events, each particle in a given event is mixed with all particles 
($\pi^{-}$ for the results presented in this paper) 
from other events, within a collection of 50 similar events. Similar events are
selected within the centrality bin
and further binned to have primary vertex {\it z} positions within 10 cm.
With the availability of high statistics data and development of new techniques,
it has become possible to have a three-dimensional
decomposition of $\vec{q}$~\cite{Bertsch:1988db,Pratt:1986cc,Chapman:1994yv},
providing better insight into the collision geometry.

The relative momentum $\vec{q}$ can be decomposed according to the
Bertsch-Pratt (also known as ``out-side-long'') convention~\cite{Abelev:2009tp}.
The relative momentum $\vec{q}$ is decomposed into the variables 
along the beam direction ($\it{q_{\rm long}}$), parallel ($\it{q_{\rm out}}$) 
to the transverse momentum of the pair
$\vec{k}_{T}$ $=$ ($\vec{p}_{\rm 1T}$ $+$ $\vec{p}_{\rm 2T}$)/2,
and perpendicular ($\it{q_{\rm side}}$) to $\it{q_{\rm long}}$ and $\it{q_{\rm out}}$.
In addition to the correlation arising from quantum statistics of two
identical particles, correlations can also arise from two-particle 
final state interactions. For identical pions, the effects
of strong interactions are negligible, but the long range Coulomb repulsion causes a
suppression of the measured correlation function at small $\vec{q}$.

In this analysis, we follow the same procedure as was used in the previous analysis
of Au+Au collisions at $\sqrt{s_{NN}}$ = 200 GeV~\cite{Adams:2004yc}. For an 
azimuthally-integrated analysis at midrapidity in the longitudinal co-moving system (LCMS), 
the correlation
function in Eq.~(\ref{eq:one}) can be decomposed as~\cite{Lisa:2005dd}: 
\begin{equation}
\label{eq:two}
C(q_{\rm out},q_{\rm side},q_{\rm long}) = ( 1 - \lambda )  + ~~~~~~~~~~~~~~~~~~~~~~~~~~~~~~~~~~~~~~~~~~
\nonumber
\end{equation}
\begin{equation}
\lambda K_{\rm coul}(q_{\rm inv})(1 + e^{-q^2_{\rm out}R^2_{\rm out} - q^2_{\rm side}R^2_{\rm side} - q^2_{\rm long}R^2_{\rm long}}),
\end{equation}
where $K_{\rm coul}$ is to a good approximation the squared nonsymmetrized Coulomb wave function
integrated over a Gaussian source 
(corresponding to the LCMS Gaussian radii 
$R_{\rm out}$, $R_{\rm side}$, $R_{\rm long}$)~\cite{lednicky}.
Assuming particle identification is perfect and the source is purely chaotic, 
$\lambda$ represents the fraction of correlated pairs emitted from the collision.

We assume a spherical Gaussian source of 5 fm~\cite{Abelev:2009tp,Adams:2004yc} 
for Au+Au collisions at $\sqrt{s_{NN}}$ = 9.2 GeV.
The first term $(1 - \lambda)$ in Eq.(\ref{eq:two}) accounts for those
pairs which do not interact or interfere.
The second term represents those pairs where both
Bose-Einstein effects and Coulomb interactions are present~\cite{Adams:2004yc}.

\bef
\includegraphics[scale=0.45]{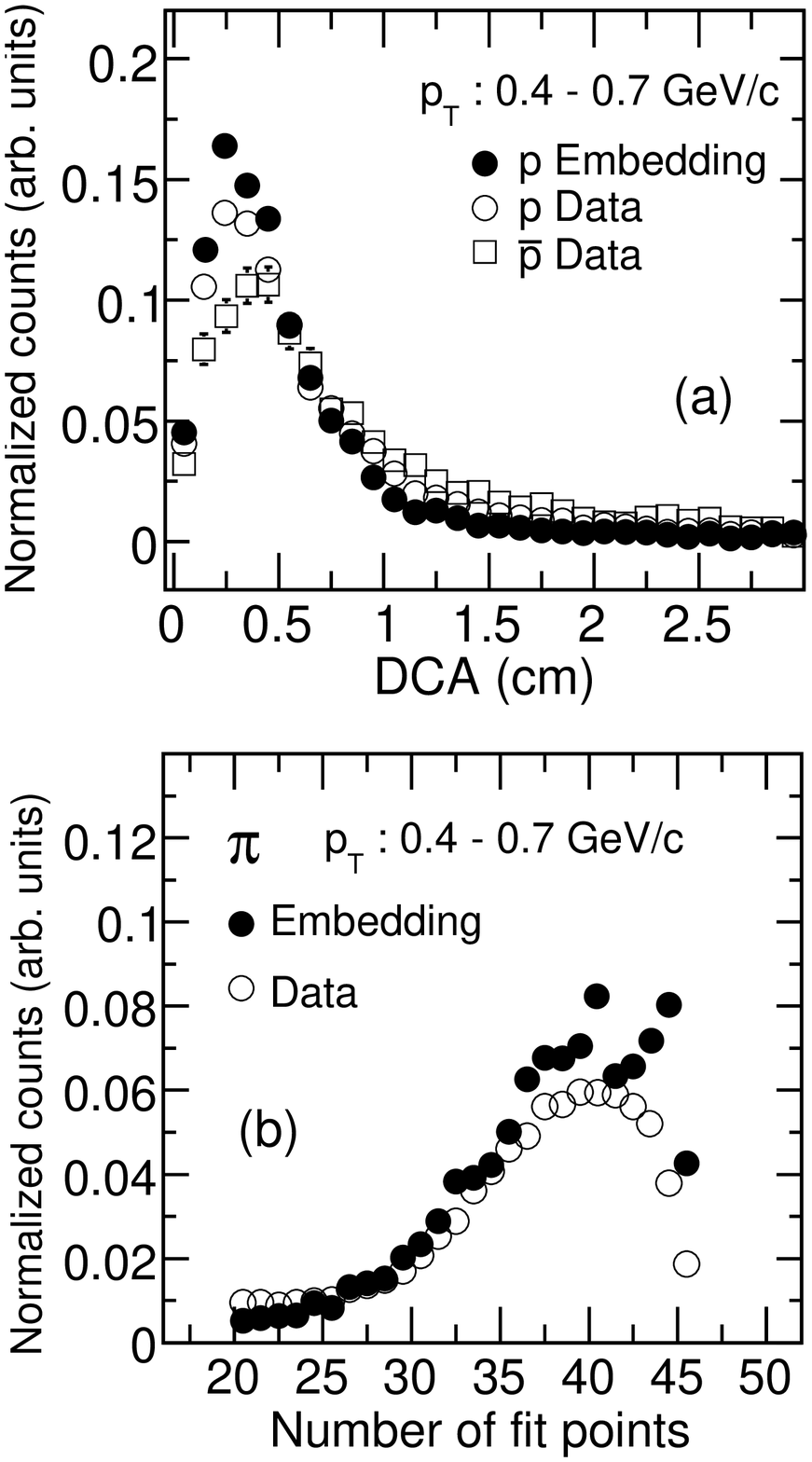}
\caption{ (a) Distribution of distance of closest approach of proton
tracks to the primary vertex. 
The embedded tracks are compared to the ones in 
real data at 0.4 $< p_{T} < 0.7$ GeV/$c$ at midrapidity in Au+Au collisions
at $\sqrt{s_{NN}}$ = 9.2 GeV. The DCA distribution
of anti-protons in a similar kinematic range is also shown for comparison.
(b) Comparison between the distributions 
of number of fit points for pions from embedding and from real data for  
0.4 $< p_{T} < 0.7$ GeV/$c$ at midrapidity in Au+Au collisions
at $\sqrt{s_{NN}}$ = 9.2 GeV. 
}
\label{dca}
\eef

\subsection{Correction factors for $\bm{p_{T}}$ spectra}

\bef
\includegraphics[scale=0.45]{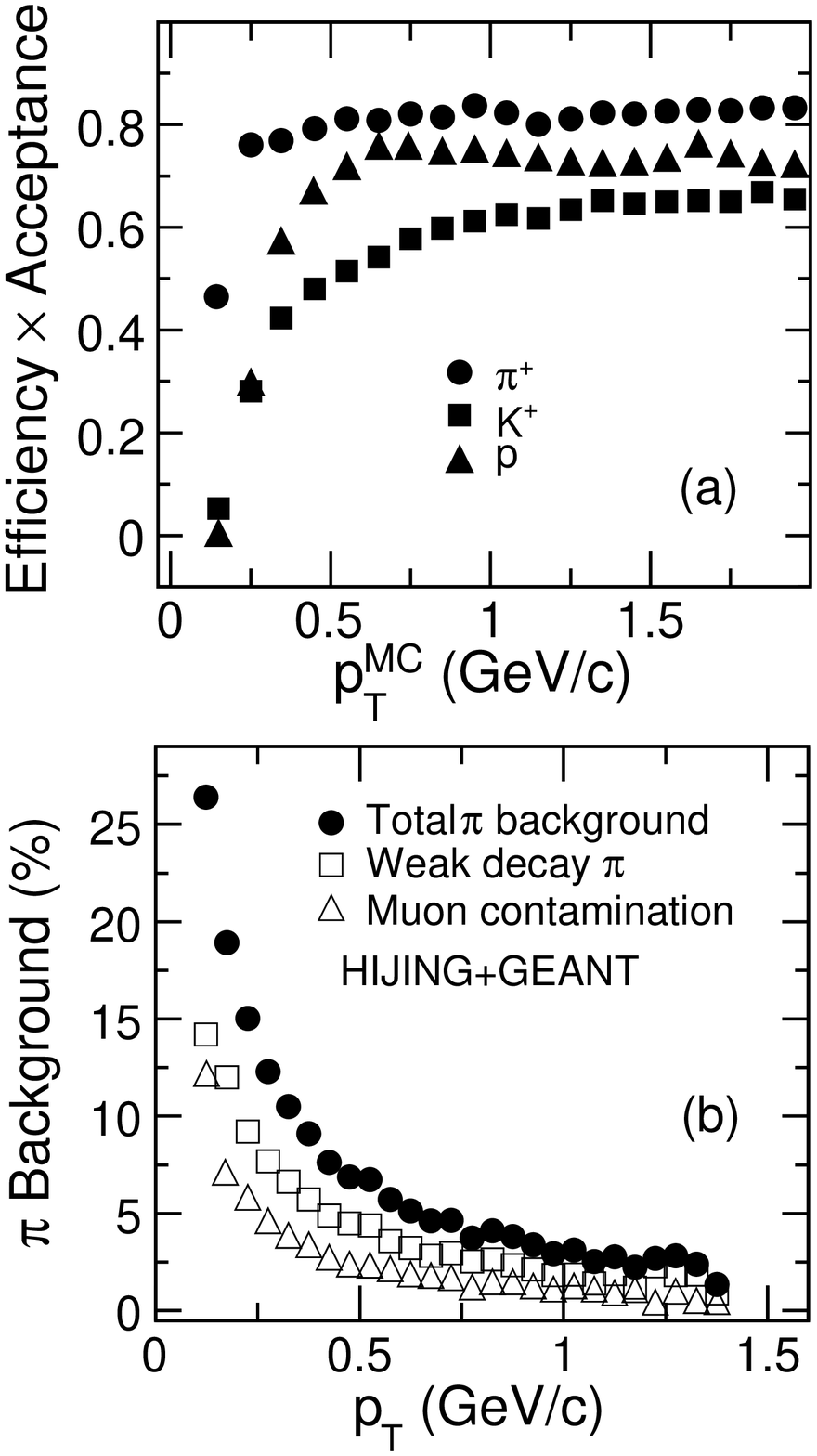}
\caption{(a) Efficiency $\times$ acceptance 
for reconstructed pions, kaons, and protons in the TPC as a function of $p_{T}$
at midrapidity in  Au+Au collisions at $\sqrt{s_{NN}}$ = 9.2 GeV. 
(b) Percentage of pion background contribution estimated from HIJING+GEANT as a function of
$p_{T}$ at midrapidity in Au+Au collisions at $\sqrt{s_{NN}}$ = 9.2 GeV.
The contributions from different sources and the total background are shown separately.
}
\label{eff}
\eef

Two major correction factors for $p_{T}$ spectra 
account for the detector acceptance and for the efficiency 
of reconstructing particle tracks. 
These are determined together by embedding 
the tracks simulated using the GEANT~\cite{geant} model of the STAR detector
into 
real events at the raw data level. One important requirement is to have a
match in the distributions of reconstructed embedded tracks and real data tracks 
for quantities reflecting track quality and used for track selection.
Figures~\ref{dca}(a) and~\ref{dca}(b) show the comparisons of DCA (for protons)
and $N_{\rm{fit}}$ 
(for pions) distributions,
respectively, 
in the range $0.4 <  p_{T} < 0.7$ GeV/$c$.
Similar 
agreement as in Fig.~\ref{dca} is observed between embedded tracks and real data 
in other measured 
$p_{T}$ ranges for all the identified hadrons presented in this paper. The ratio 
of the distribution of reconstructed and original Monte Carlo tracks
as a function of $p_{T}$ gives the acceptance $\times$ efficiency correction 
factor as a function of $p_{T}$ for the rapidity interval
studied. The typical efficiency $\times$ acceptance factors in 0--60\% central
collisions for pions, kaons and protons at midrapidity ($|y| < 0.5$) are shown in 
Fig.~\ref{eff}(a). The raw yields are corrected by these factors to obtain the 
final $p_{T}$ spectra.

The STAR experiment has previously observed that proton yields had significant 
contamination from secondary protons, due to interactions of
energetic particles produced in collisions with detector materials. 
As these secondary protons are produced away from the primary interaction point, they 
appear as a long tail in the DCA distribution of protons. 
A comparison between shapes of DCA distributions of protons and anti-protons
(which do not have such sources of background) was used 
in STAR to estimate the background contribution to the proton yield~\cite{STARPID, starprc70}.
This feature was found to be more pronounced at lower $p_{T}$. In this 
test run, it is observed that the DCA distribution for protons does not exhibit
a long tail, and that for all the $p_{T}$ ranges studied, its shape is similar to
that for anti-protons (Fig.~\ref{dca}(a), 
distributions normalized to the same number of total counts). 
This lack of secondary protons for Au+Au collisions
at $\sqrt{s_{NN}} =$ 9.2 GeV 
could be due to the experimental configuration in the year 2008 with reduced amount 
of material in 
front of the STAR TPC, and due to the relatively
small number of energetic particles produced in the interactions compared to collisions
at higher energies of $\sqrt{s_{NN}} =$ 62.4 and 200 GeV. 
No corrections for secondary proton background are applied for the present 
analysis at $\sqrt{s_{NN}}$ = 9.2 GeV.

The charged pion spectra are corrected for feed-down from
weak-decays, muon contamination, and background pions
produced in the detector materials. These corrections are
obtained from Monte Carlo simulations of HIJING events at $\sqrt{s_{NN}}$ = 9.2 GeV, 
with the STAR geometry for year 2008 and a realistic description of the detector
response used in GEANT. The simulated events are reconstructed
in the same way as the real data. The weak-decay daughter
pions are mainly from $K^{0}_{S}$, and are identified by the parent particle 
information accessible from the simulation. 
The muons from pion decay can 
be misidentified as primordial pions 
due to their similar masses.
This contamination is obtained from Monte Carlo simulations by
identifying the decay, which is accessible in the
simulation. 
The weak-decay pion background
and muon contamination obtained from the simulation are shown in Fig.~\ref{eff}(b), 
as a function of simulated pion $p_{T}$ for 0--60\% central Au+Au collisions at 
$\sqrt{s_{NN}}$ = 9.2 GeV.  
The final pion spectra are corrected for this background effect. 

The low momentum particles lose energy while traversing the detector material.
The track reconstruction algorithm takes into account the
Coulomb scattering and energy loss, assuming the
pion mass for each particle. Therefore, a correction for the
energy loss by heavier particles ($K^{\pm}$, $p$ and $\bar{p}$) is needed. 
This correction is obtained from embedding Monte Carlo simulations. The largest 
change in reconstructed $p_{T}$ is found to be $\sim$20 MeV/$c$ at $p_{T} = $ 200 MeV/$c$. 
For all results presented in this paper, the track $p_{T}$ is corrected for this 
energy loss effect.

\subsection{Systematic errors}

\begin{table}
\caption{Sources of systematic errors on yields of various produced hadrons.
See section II H for more details.
\label{table3}}
\vspace{0.15cm}
\begin{tabular}{c|c|c|c|c|c|c}
\hline
Hadron &$V_{z}$ & cuts & $y$ & correction & PID & extrapolation \\
\hline
$\pi$ & 3\%  &  3.2\%   & 2\%   & 5\% & 5\%  & 3\%\\
$K$     & 3\%  &  6.2\%   & 2\%   & 5\% & 10\% & 8\%\\
$p$     & 3\%  &  5.4\%   & 10\%  & 5\% & 4\%  & 15\%\\
\hline
\end{tabular}
\end{table}

Systematic uncertainties on the spectra are estimated by varying 
cuts, and by assessing the purity of identified hadron sample
from $dE/dx$ measurements. In addition, the Gaussian
fit ranges are varied to estimate the systematic uncertainty
on the extracted raw spectra. 
The point-to-point
systematic errors are quoted in figure captions. The statistical and
systematic errors are added in quadrature and 
plotted for most of the results unless otherwise specified.
For integrated particle yields, 
extrapolating yields to unmeasured regions in $p_{T}$ is an additional source of
systematic error. These are estimated by comparing 
the extrapolations using different fit functions to the $p_{T}$ spectra. 
The detailed procedure is described in Ref.~\cite{STARPID}. A summary of various sources
of systematic errors on the identified hadron yields for 0--60\% centrality 
in Au+Au collisions at $\sqrt{s_{NN}}$ = 9.2 GeV is given 
in Table~\ref{table3}. 
The column titled ``$V_{z}$'' in Table~\ref{table3} represents the systematic errors
obtained by varying the $V_{z}$ range in the analysis, ``cuts'' lists systematic 
errors due to variation of DCA and $N_{\rm{fit}}$ cut values, ``$y$'' represents the systematic
effect on yields due to a variation in rapidity range from $\pm$ 0.5 to $\pm$ 0.2,
``correction'' includes the contribution to systematic errors from 
track reconstruction efficiency and acceptance estimates,
``PID'' represents the systematic errors associated with particle
identification (obtained by varying the $dE/dx$ cuts and the range of Gaussian fits to
normalized $dE/dx$ distributions), and 
``extrapolation'' refers to the contribution of systematic 
errors from the different fit functions used for obtaining yields in
unmeasured $p_{T}$ ranges. 
In addition, the systematic error arising due to the
pion background estimation (discussed in the previous subsection) is also calculated.
It is of the order of 6\%.
The total systematic errors are of the order of 11\%, 16\%, and 20\% for pion,
kaon, and proton yields respectively.

The systematic errors in the directed flow analysis are obtained,
(a) by exploiting the symmetry in the measurements
for forward and backward regions 
with respect to $\eta = 0$, and
(b) by comparing $v_{1}$ calculated from different methods with  
various sensitivities to non-flow effects~\cite{flow1}.
In (a), we average $v_1$ from the mixed harmonics  
method ($v_{1}\{ \rm {EP_{1}}, \rm {EP_{2}}\}$) and from the two standard methods 
($v_{1}\{ \rm{EP_{1}}, \rm{FTPC}\}$  and $v_{1}\{ \rm{EP_{1}}, \rm{BBC}\}$), 
as discussed in section II E, and take the difference between the magnitude of $v_1$ 
in the forward and backward region as the systematic error due to the  
unbalanced detector response. 
We report
an absolute error of $\sim$7.8\% in the FTPC range (2.5 $<|\eta|<$ 4.2), 
and negligible error in the TPC range.
In (b), we average the magnitude of $v_1$ in the forward and backward  
region, and take the maximum difference between results from the three  
methods as the systematic uncertainty. 
An absolute error of $\sim$10\% is found 
for the FTPC range, and $\sim$50\% relative error for  
the TPC range. The $v_{1}\{ \rm{EP_{1}}, \rm{BBC}\}$ method in the
TPC range ($|\eta| <$ 1.3) is more reliable compared to the other two  
methods. This is due
to the large $\eta$ gap between the BBC and TPC detectors, which helps  
subtract the non-flow
effect. 
The $\eta$ gap between the BBC and the TPC is up to 2.6 units, while it is  
only 1.3 units
between the FTPC and the TPC.
The total absolute systematic error on the $v_{1}$ estimate is calculated
as the quadrature sum of components (a) and (b), which is $\sim$10\% 
(absolute error),  for the FTPC range and $\sim$50\% (relative  
value) for the TPC range.

The systematic errors on the elliptic flow parameter are evaluated by varying the 
event vertex selection along the beam direction, varying the DCA cut value, and by using the $\eta$ 
sub-event method. The total systematic error on $v_{2}$ is
approximately 10\%.

For the pion interferometry analysis, we study the following sources of systematic error:  
track merging, track splitting, size of the source used for Coulomb correction, particle 
identification, and 
pair acceptance for pions of opposite charges.
The estimated systematic errors are less than 10$\%$ for 
all radii in the 0--60$\%$ centrality bin for $150 < k_{T} < 250$ MeV/$c$,
similar to those in Refs.~\cite{Abelev:2009tp,Adams:2004yc}.

\bef
\includegraphics[scale=0.45]{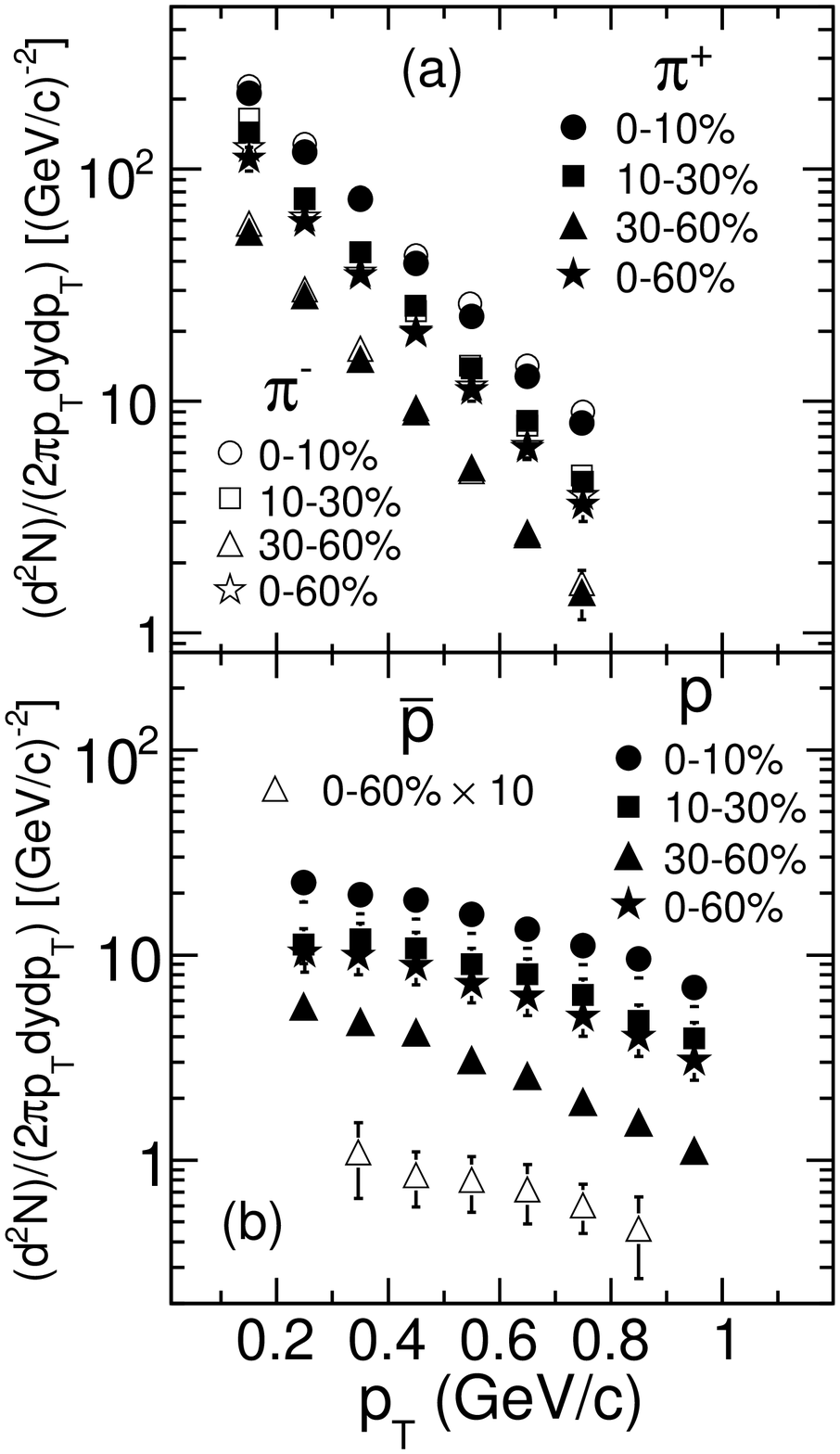}
\caption{Transverse momentum spectra for (a) charged pions and (b) protons 
at midrapidity ($|y|<0.5$) in Au+Au collisions at $\sqrt{s_{NN}}$ = 9.2 GeV 
for various centralities. The distributions for anti-protons were measured 
in this limited statistics data only for 0--60\% centrality. The anti-proton yield  
shown in the figure is multiplied by a factor of 10. 
The errors shown are statistical and systematic errors (discussed in section II H)
added in quadrature.}
\label{ptspectra_pip}
\eef

\bef
\includegraphics[scale=0.45]{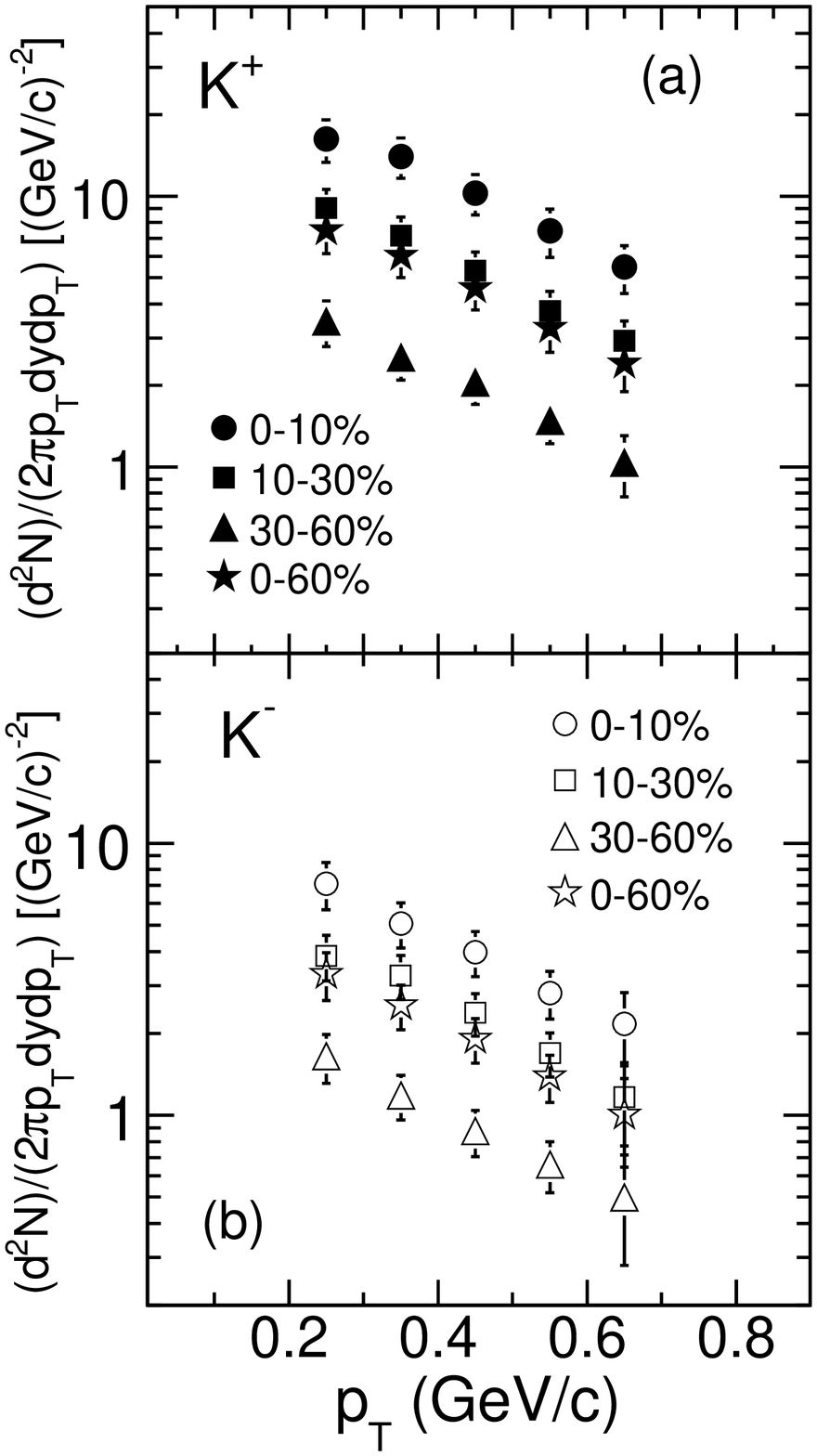}
\caption{Transverse momentum spectra for (a) positive kaons and 
(b) negative kaons at midrapidity ($|y|<0.5$) in Au+Au collisions at 
$\sqrt{s_{NN}} =$ 9.2 GeV 
for various centralities. 
The errors shown are statistical and systematic errors (discussed in section II H)
added in quadrature.}
\label{ptspectra_k}
\eef

\section{Results}

\subsection{Transverse momentum spectra}

Figures~\ref{ptspectra_pip}(a) and~\ref{ptspectra_pip}(b) show the transverse momentum spectra 
for $\pi^{\pm}$ and $p$ ($\bar{p}$), respectively.
Figure~\ref{ptspectra_k}(a) and~\ref{ptspectra_k}(b) show the spectra for $K^{+}$ and
$K^{-}$, respectively, in Au+Au collisions at $\sqrt{s_{NN}}$ = 9.2 GeV. The results 
are shown for 
the collision centrality classes of 0--10\%, 10--30\%, 30--60\%, and 0--60\%. The $\bar{p}$ 
spectrum is shown only for 0--60\% centrality and the yields are multiplied by a factor 
of 10 for visibility.
The inverse slopes of
the identified hadron spectra follow the order $\pi$ $<$ $K$ $<$ $p$. 
An exponential fit  to the  $p_{T}$
spectra of 
$\pi^{+}$, $K^{+}$, and $p$ yields inverse slopes of 180 $\pm$ 7 MeV, 
360 $\pm$ 7 MeV and 616 $\pm$ 11 MeV respectively. The errors on the inverse slopes 
are statistical. The spectra can be further characterized by looking at 
the $dN/dy$ and $\langle p_{T} \rangle$ or 
$\langle m_{T} \rangle - m$ for the produced hadrons, where $m$ is the mass of the hadron and 
$m_{T}$ = $\sqrt{m^{2} + p_{T}^{2}}$ is its transverse mass. 
Those observables are discussed in the 
following sections.

\subsection{Centrality dependence of particle production}

\bef
\includegraphics[scale=0.45]{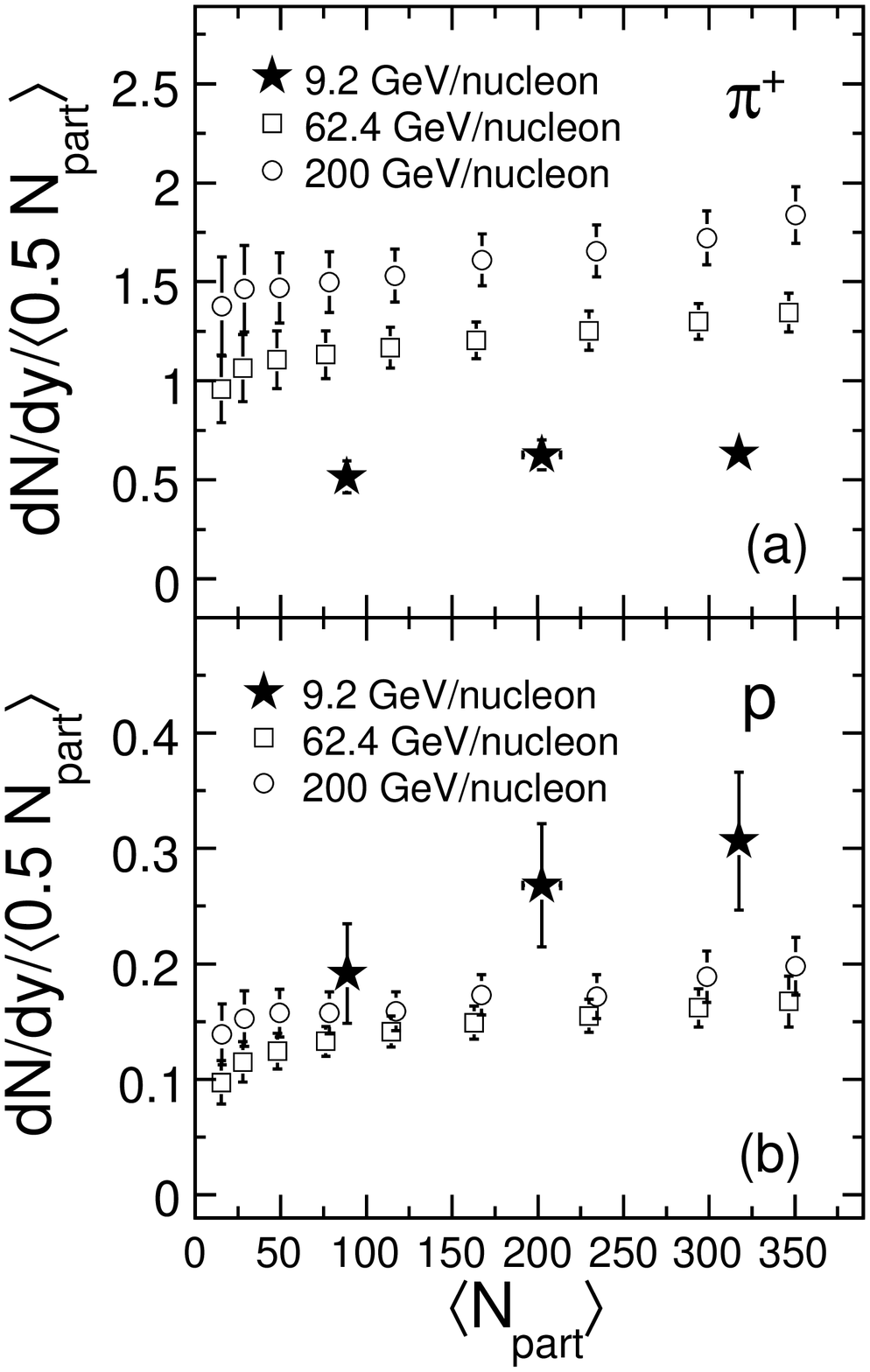}
\caption{$dN/dy$ of (a) $\pi^{+}$ and (b) $p$, normalized by $\langle N_{\mathrm {part}} \rangle / 2$, 
for Au+Au collisions at $\sqrt{s_{NN}}$ = 9.2 GeV, plotted as a function 
of $\langle N_{\mathrm {part}} \rangle$. The lower energy results are compared to corresponding
results for Au+Au collisions at $\sqrt{s_{NN}}$ = 62.4 and 200 GeV~\cite{STARPID, starprl92}. 
The errors shown are the quadrature sum of statistical and systematic uncertainties.
The systematic errors on pion and proton yields for $\sqrt{s_{NN}}$ = 9.2 GeV data
are $\sim$ 12\% and $\sim$ 20\%, respectively, for all the collision centralities studied.}
\label{dndycent_pip}
\eef

\bef
\includegraphics[scale=0.45]{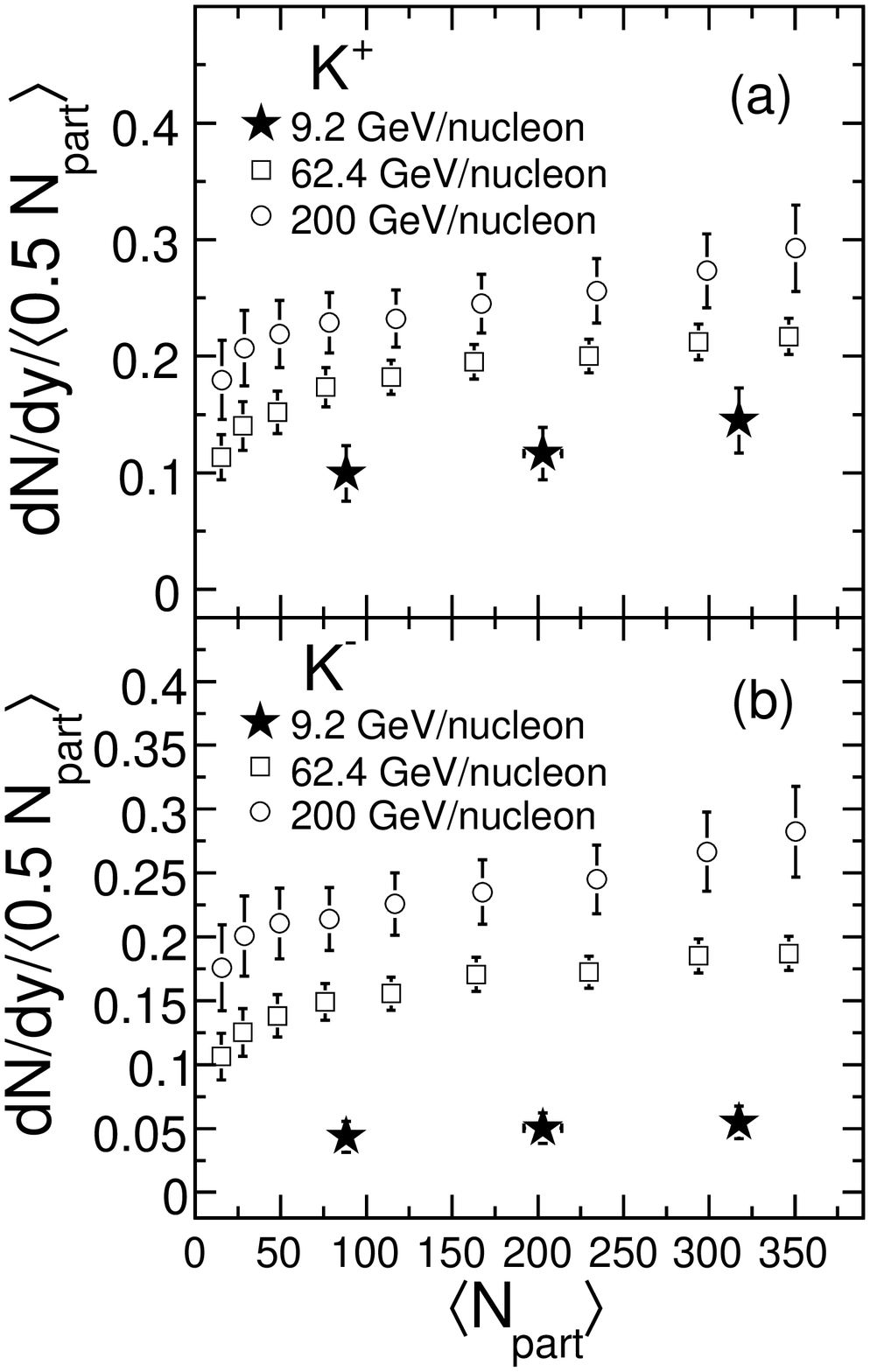}
\caption{$dN/dy$ of (a) $K^{+}$ and (b) $K^{-}$, normalized by $\langle N_{\mathrm {part}} \rangle / 2$ 
for Au+Au collisions at $\sqrt{s_{NN}}$ = 9.2 GeV, plotted as a function 
of $\langle N_{\mathrm {part}} \rangle$. The lower energy results are compared to corresponding
results for Au+Au collisions at $\sqrt{s_{NN}}$ = 62.4 and 200 GeV~\cite{starraa,STARPID,starprl92}.
The errors shown are the quadrature sum of statistical and systematic uncertainties.
The systematic errors on $K^{+}$ and $K^{-}$ yields for $\sqrt{s_{NN}}$ = 9.2 GeV data
are similar, about 18\%  for all the collision centralities studied.
}
\label{dndycent_k}
\eef

Figures~\ref{dndycent_pip} and~\ref{dndycent_k} show the comparison of collision centrality 
dependence of 
$dN/dy$ of $\pi^{+}$, $K^{\pm}$, and $p$, normalized by $\langle N_{\mathrm {part}} \rangle / 2$,
between new results at $\sqrt{s_{NN}}$ = 9.2 GeV and previously published results 
at $\sqrt{s_{NN}}$ = 62.4 and 200 GeV from the STAR experiment~\cite{starraa,STARPID,starprl92}. 
The yields of charged pions and kaons decrease with decreasing collision energy.
The collision centrality
dependence within the limited centrality region studied for the new results is 
similar to that at higher beam energies. 
For protons, the yield is larger in central Au+Au collisions at $\sqrt{s_{NN}} =$ 9.2 GeV 
compared to corresponding results at $\sqrt{s_{NN}} =$ 62.4 and 
200 GeV~\cite{starraa,STARPID,starprl92}. 
For the most peripheral
collisions, the yields are comparable within errors to corresponding yields at higher beam
energies. The increase in proton yield 
with the increasing collision 
centrality is due to large net-proton ($p - \bar{p}$) density at midrapidity in the lower 
collision energies.

The inclusive $dN_{\rm ch}/d\eta$ (sum of contributions from
$\pi^{\pm}$, $K^{\pm}$, and $p$ $(\bar{p})$ found by redoing 
the analysis binned in $\eta$ instead of rapidity)
at midrapidity 
for various collision centralities are
given in Table~\ref{dnchdeta} along with the statistical and systematic errors
for Au+Au collisions at $\sqrt{s_{NN}}$ = 9.2 GeV.
\begin{table}
\caption{ Centrality dependence of $dN_{\mathrm {ch}}/d\eta$ at midrapidity in 
Au+Au collisions at $\sqrt{s_{NN}}$ = 9.2 GeV.
\label{dnchdeta}}
\vspace{0.15cm}
\begin{tabular}{c|c|c|c}
\hline
\% cross section& $dN_{\mathrm {ch}}/d\eta$ &~stat. error &~sys. error\\ [0.2mm]
\hline
0--10  &  229   & 25  & 62\\
10--30 &  133   & 15  & 36\\
30--60 &  48    & 5   & 13\\
\hline
\end{tabular}
\end{table}

\bef
\begin{center}
\includegraphics[scale=0.39]{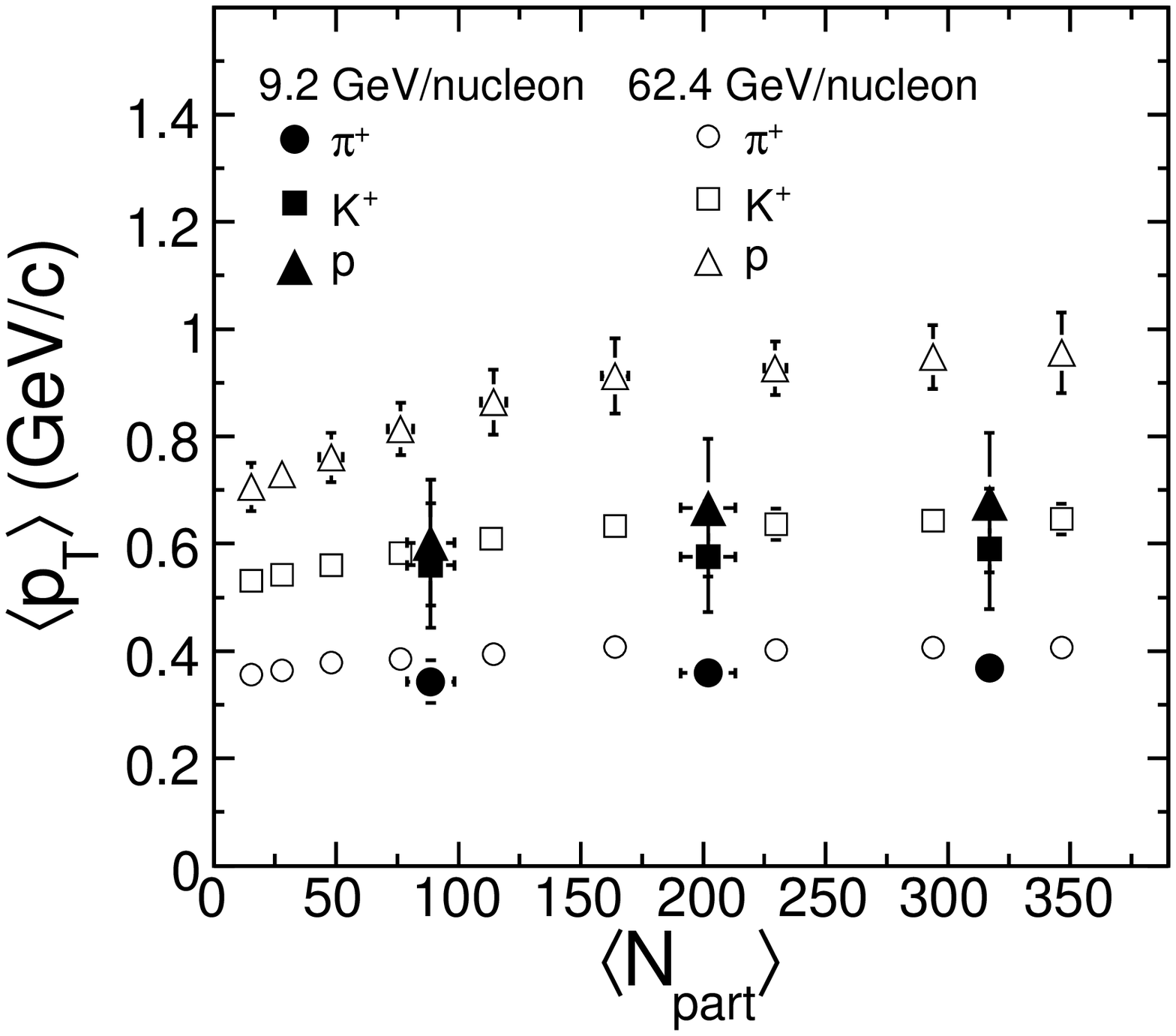}
\caption{$\langle p_{T} \rangle$ for $\pi^{+}$, $K^{+}$, and $p$ plotted 
as a function of $\langle N_{\mathrm {part}} \rangle$ for Au+Au collisions at
$\sqrt{s_{NN}} =$ 9.2 GeV and compared to corresponding results at $\sqrt{s_{NN}} =$ 62.4 
GeV~\cite{starraa,STARPID,starprl92}.
The errors shown are the quadrature sum of statistical and systematic uncertainties.
The systematic errors for pions, kaons, and protons for $\sqrt{s_{NN}}$ = 9.2 GeV 
are $\sim$  12\%, 18\%, and 21\%  
respectively, and similar for all the collision centralities studied.
} 
\label{meanpt62}
\end{center}
\eef
\bef
\begin{center}%
\includegraphics[scale=0.39]{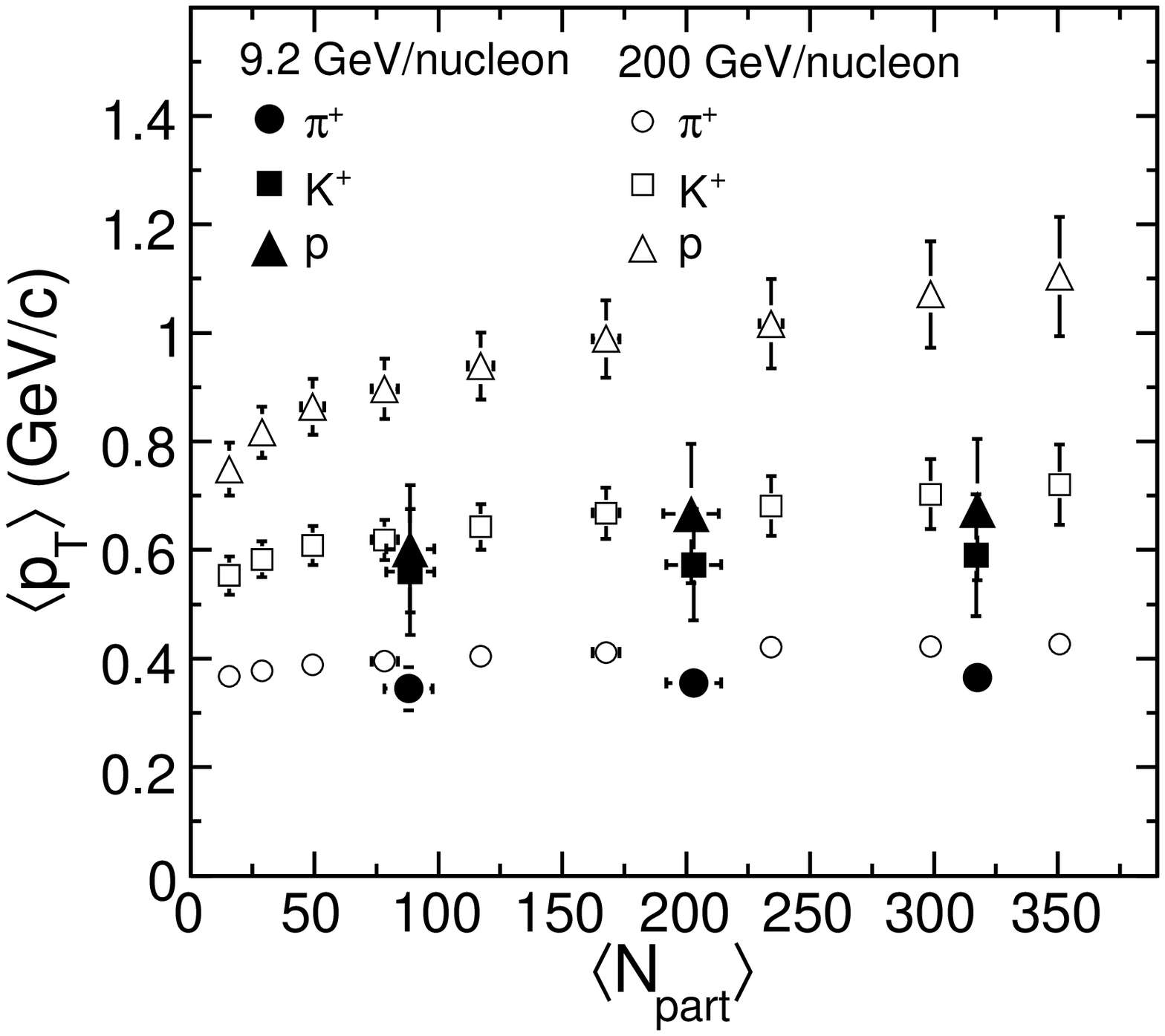}
\caption{$\langle p_{T} \rangle$ for $\pi^{+}$, $K^{+}$ and $p$ plotted
as a function of $\langle N_{\mathrm {part}} \rangle$ for Au+Au collisions at
$\sqrt{s_{NN}}$ = 9.2 GeV and compared to corresponding results at $\sqrt{s_{NN}} =$ 
200 GeV~\cite{starraa,STARPID, starprl92}.
The errors shown are the quadrature sum of statistical and systematic uncertainties.
The systematic errors for pions, kaons, and protons for $\sqrt{s_{NN}}$ = 9.2 GeV 
are $\sim$  12\%, 18\%, and 21\%  
respectively, and similar for all the collision centralities studied.
}
\label{meanpt200}
\end{center}
\eef

Figures~\ref{meanpt62} and~\ref{meanpt200} show the comparison of $\langle p_{T} \rangle$
as a function of $\langle N_{\mathrm {part}} \rangle$ for $\pi^{+}$, $K^{+}$, and $p$ from 
Au+Au collisions  at $\sqrt{s_{NN}}$ = 9.2 GeV with those from collisions 
at  $\sqrt{s_{NN}}$ = 62.4 and 
200 GeV~\cite{starraa,STARPID,starprl92}.
For the collision centralities studied, the dependencies of $\langle p_{T} \rangle$
on $\langle N_{\mathrm {part}} \rangle$ at $\sqrt{s_{NN}}$ = 9.2 GeV
are similar to those at $\sqrt{s_{NN}} = $ 62.4 and 200 GeV. An increase in 
$\langle p_{T} \rangle$ with increasing hadron mass is observed 
at 
$\sqrt{s_{NN}}$ = 9.2 GeV. A similar dependence is also observed for $\sqrt{s_{NN}} =$ 
62.4 and 200 GeV.
However, the differences in $\langle p_{T} \rangle$ between protons and kaons are much 
smaller compared to the observations at higher beam energies. The mass dependence of 
$\langle p_{T} \rangle$
reflects collective expansion in the radial direction. The smaller difference between 
$\langle p_{T} \rangle$ 
of protons and kaons at 
$\sqrt{s_{NN}} = $ 9.2 GeV indicates that the average collective velocity in the radial 
direction is smaller at that energy.

\bef
\includegraphics[scale=0.45]{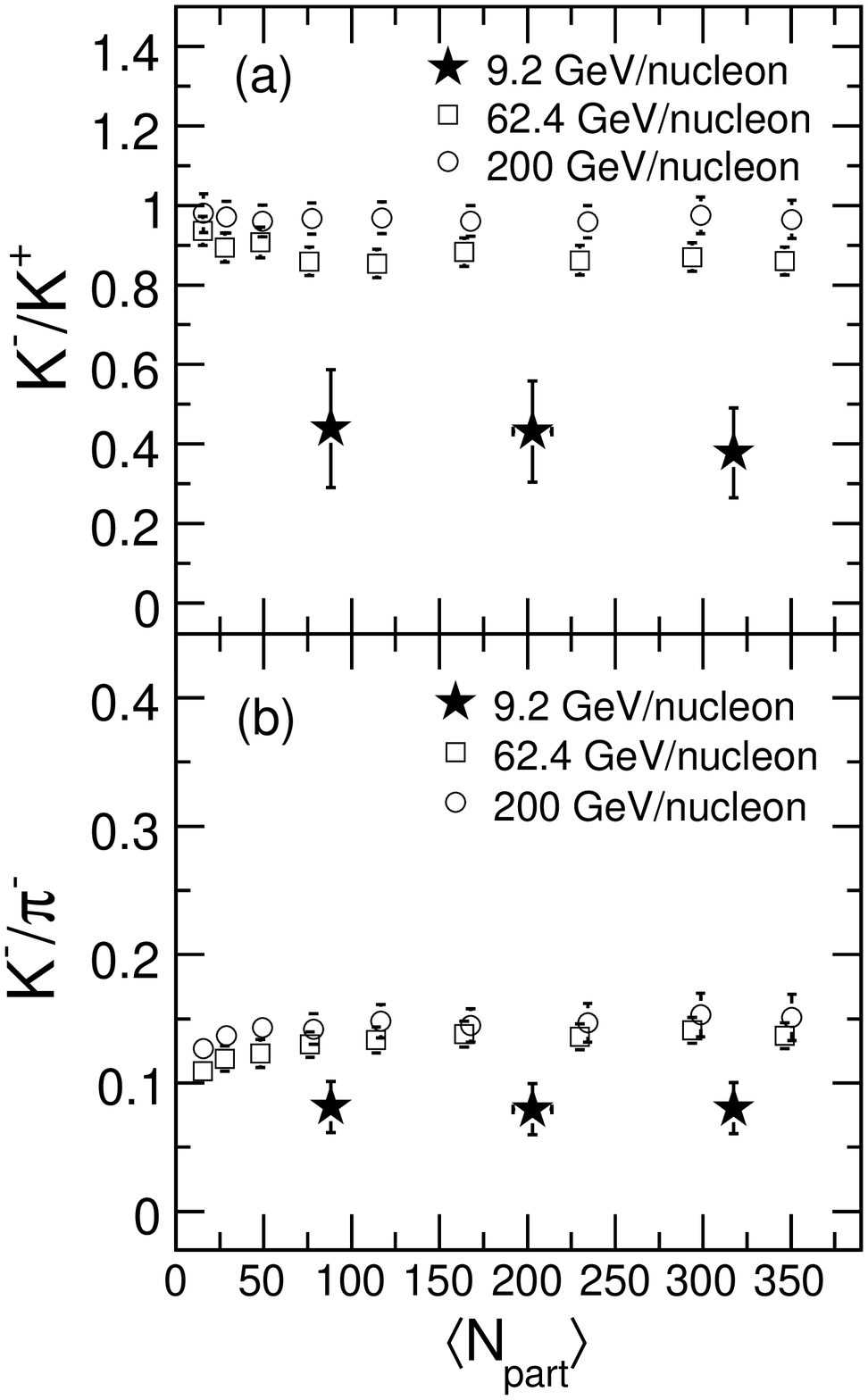}
\caption{Variation of (a) $K^{-}$/$K^{+}$ and (b) $K^{-}$/$\pi^{-}$
ratios as a function of $\langle N_{\mathrm {part}} \rangle$
for Au+Au collisions at $\sqrt{s_{NN}}$ = 9.2 GeV. For comparison we also
show the corresponding results from Au+Au collisions at $\sqrt{s_{NN}}$ = 62.4 and 
200 GeV~\cite{STARPID,starprl92}.
The errors shown are the quadrature sum of statistical and systematic uncertainties.
The systematic errors for $K^{-}$/$K^{+}$ and $K^{-}$/$\pi^{-}$ for $\sqrt{s_{NN}}$ = 9.2 GeV data 
are $\sim$  25\% and 22\%,  respectively, and similar for all the collision centralities studied.
}
\label{centratio}
\eef

\bef
\includegraphics[scale=0.45]{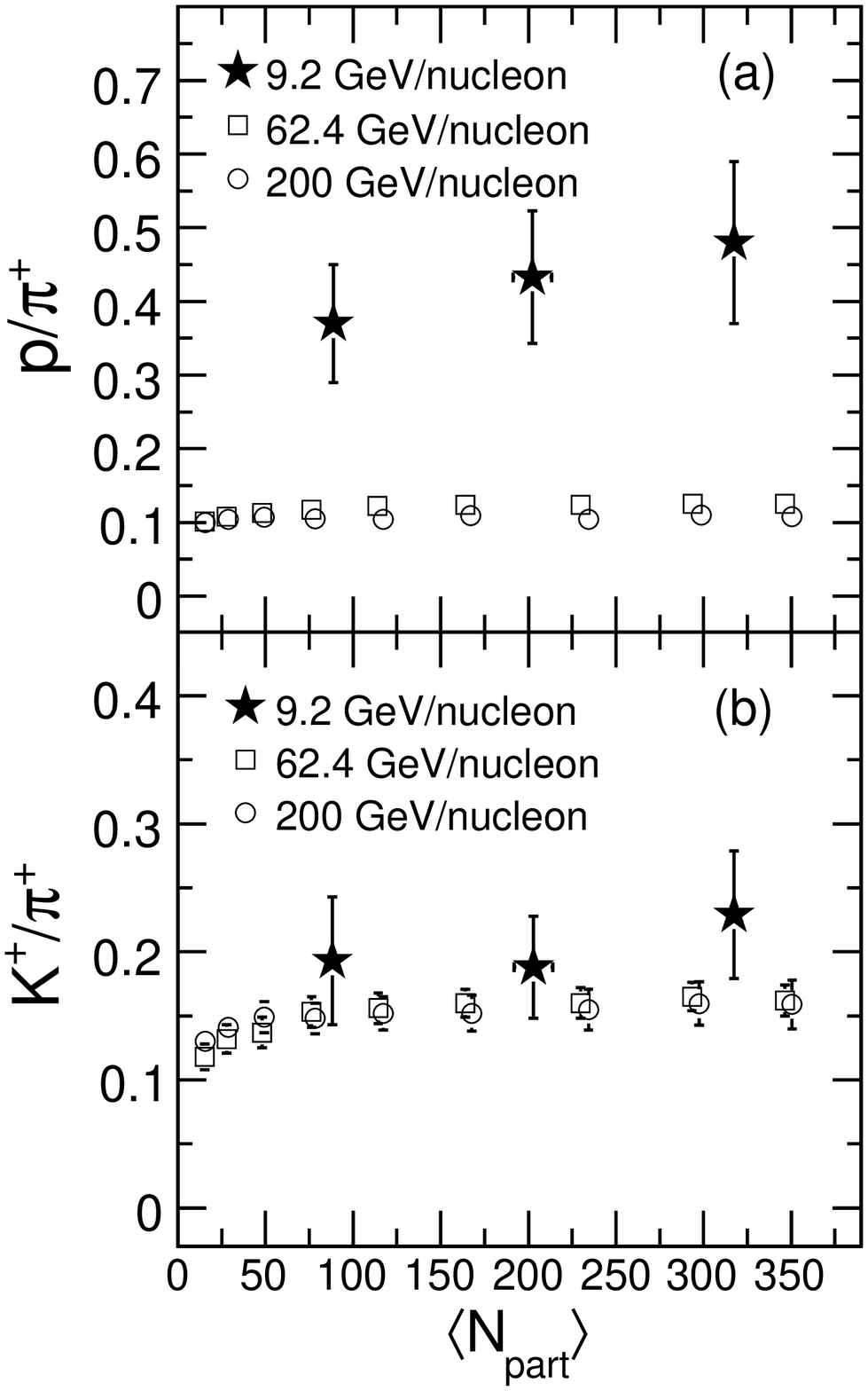}
\caption{Variation of (a) $p$/$\pi^{+}$ and (b) $K^{+}$/$\pi^{+}$ ratios as 
a function of $\langle N_{\mathrm {part}} \rangle$
for Au+Au collisions at $\sqrt{s_{NN}}$ = 9.2 GeV. For comparison we also
show the corresponding results from Au+Au collisions at $\sqrt{s_{NN}}$ = 62.4 
and 200 GeV~\cite{starraa,STARPID,starprl92}. 
The errors shown are the quadrature sum of statistical and systematic uncertainties.
The systematic errors for $p$/$\pi^{+}$  and $K^{+}$/$\pi^{+}$ for $\sqrt{s_{NN}}$ = 9.2 GeV data 
are $\sim$  25\% 
and 22\%,  respectively, and similar for all the collision centralities studied.
}
\label{centratio1}
\eef

Figures~\ref{centratio} and~\ref{centratio1} show the various particle 
ratios ($K^{-}$/$K^{+}$, $K^{-}$/$\pi^{-}$, $p$/$\pi^{+}$,
and $K^{+}$/$\pi^{+}$) as a function of collision centrality expressed as 
$\langle N_{\mathrm {part}} \rangle$
for Au+Au collisions at $\sqrt{s_{NN}} = $ 9.2 GeV. 
Corresponding results from Au+Au collisions at $\sqrt{s_{NN}} = $ 62.4 and 
200 GeV~\cite{starraa,STARPID, starprl92} are also shown. 
The $\pi^{-}$/$\pi^{+}$ ratio is close to unity and is not shown. 
Due to low event statistics and the low yield of anti-protons,
the centrality dependence of the $\bar{p}$/$p$ ratio for $\sqrt{s_{NN}} = $ 9.2 GeV 
collisions could not be extracted. 

The $K^{-}$/$K^{+}$ and
$K^{-}$/$\pi^{-}$ ratios are lower at $\sqrt{s_{NN}} =$ 9.2 GeV 
compared to those 
at $\sqrt{s_{NN}}$ = 62.4 and 200 GeV. 
In the case of $K^{+}$/$\pi^{+}$, there is less variation between 9.2 GeV and the highest
RHIC energies than in case of the other particle ratios discussed above.
This reflects an interplay
between the decreasing importance of associated production and
an increasing contribution from pair production of kaons with
increasing collision energy. 
Associated production refers to reactions 
such as $NN \rightarrow KYN$ and $\pi N \rightarrow KY$, where $N$ is a nucleon and $Y$ a hyperon.
The $p$/$\pi^{+}$ ratio is larger at $\sqrt{s_{NN}} = $ 9.2 GeV 
than at $\sqrt{s_{NN}} = $ 62.4 and 200 GeV for all collision centralities studied. As 
discussed above, this 
is a consequence of higher net-proton density at midrapidity for the collisions at 
$\sqrt{s_{NN}} = $ 9.2 GeV compared to those at $\sqrt{s_{NN}} = $ 62.4 and 200 GeV.

\bef
\includegraphics[scale=0.39]{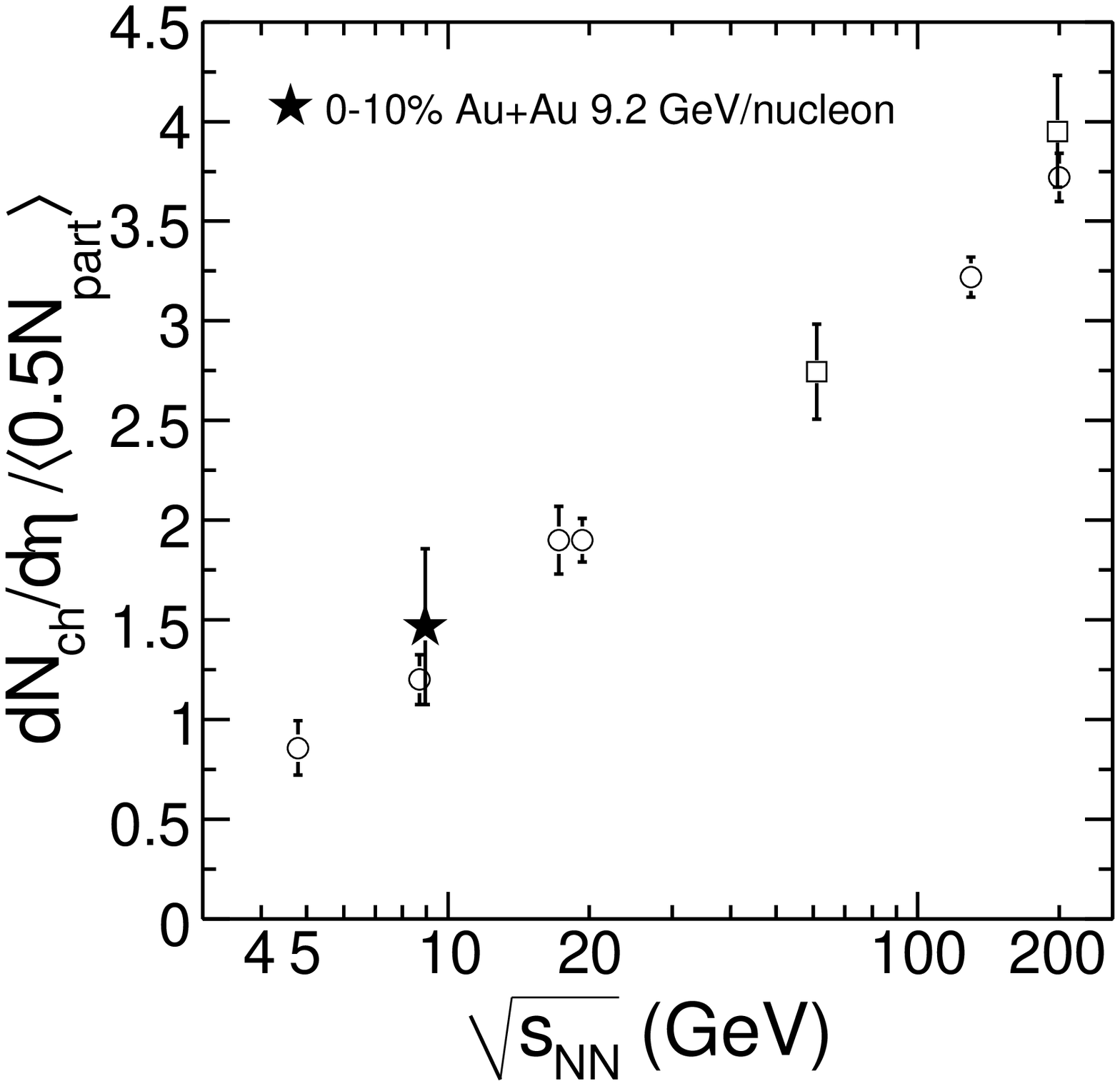}
\caption{The midrapidity $dN_{\mathrm {ch}}/d\eta$ normalized by 
$\langle N_{\mathrm {part}} \rangle$/2 as a function of $\sqrt{s_{NN}}$.
Au+Au collisions at $\sqrt{s_{NN}}$ = 9.2 GeV are compared to previous 
results from AGS~\cite{ags}, SPS~\cite{sps}, and RHIC~\cite{STARPID,PHENIXnch}. 
The error shown are the quadrature sum of statistical and systematic uncertainties.
The systematic error for $\sqrt{s_{NN}}$ = 9.2 GeV data is 39\%.}
\label{dndetaen}
\eef
\bef
\includegraphics[scale=0.45]{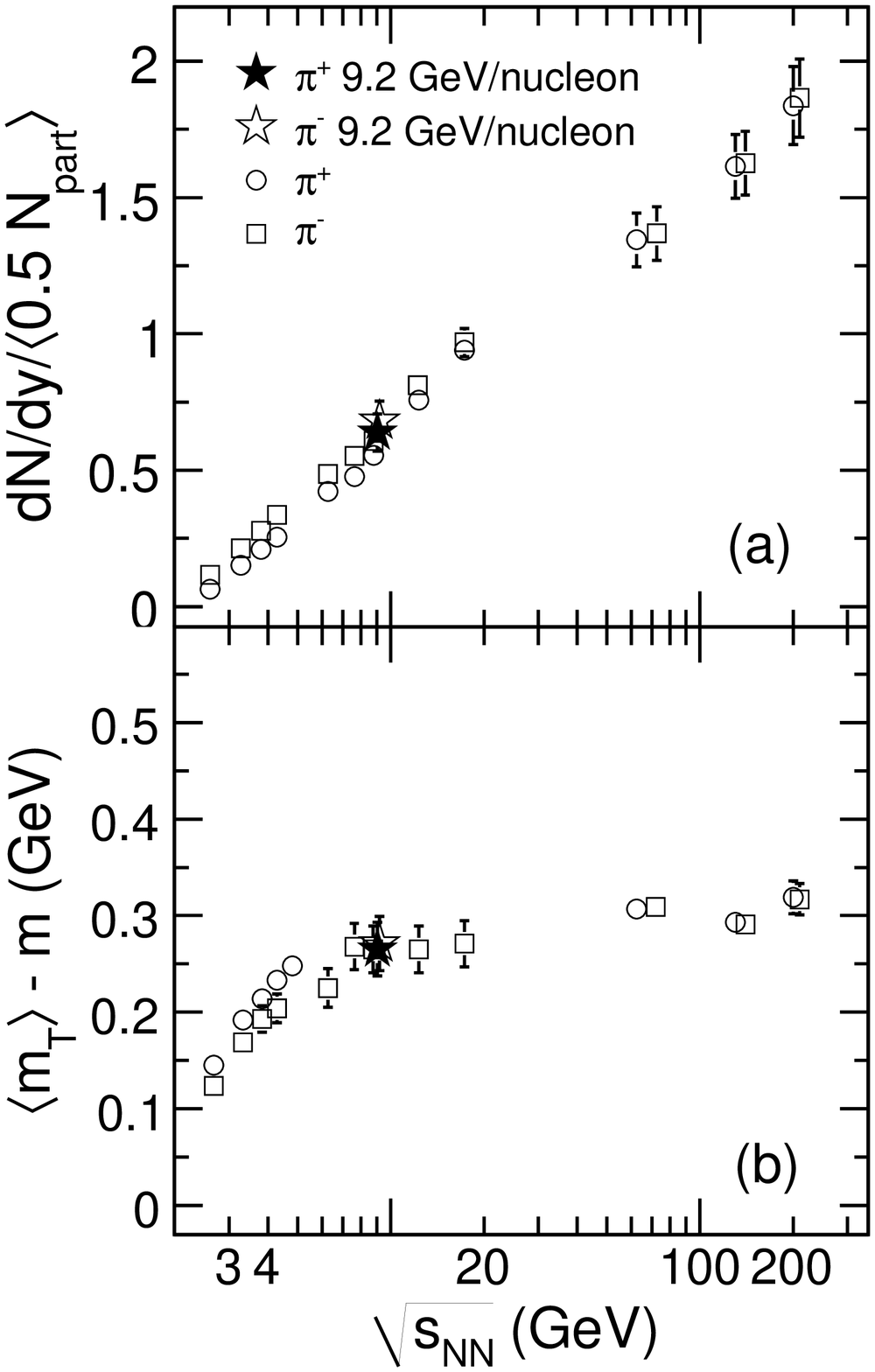}
\caption{(a) $dN/dy$ normalized by 
$\langle N_{\mathrm {part}} \rangle$/2 and 
(b) $\langle m_{T} \rangle - m$ of $\pi^{\pm}$, in 0--10\% central 
Au+Au collisions for $\sqrt{s_{NN}} = $ 9.2 GeV compared to previous 
results from AGS~\cite{ags}, SPS~\cite{sps}, and RHIC~\cite{STARPID}. 
The errors shown are the quadrature sum of statistical and systematic
uncertainties.
}
\label{energydndymt_pi}
\eef
\bef
\includegraphics[scale=0.45]{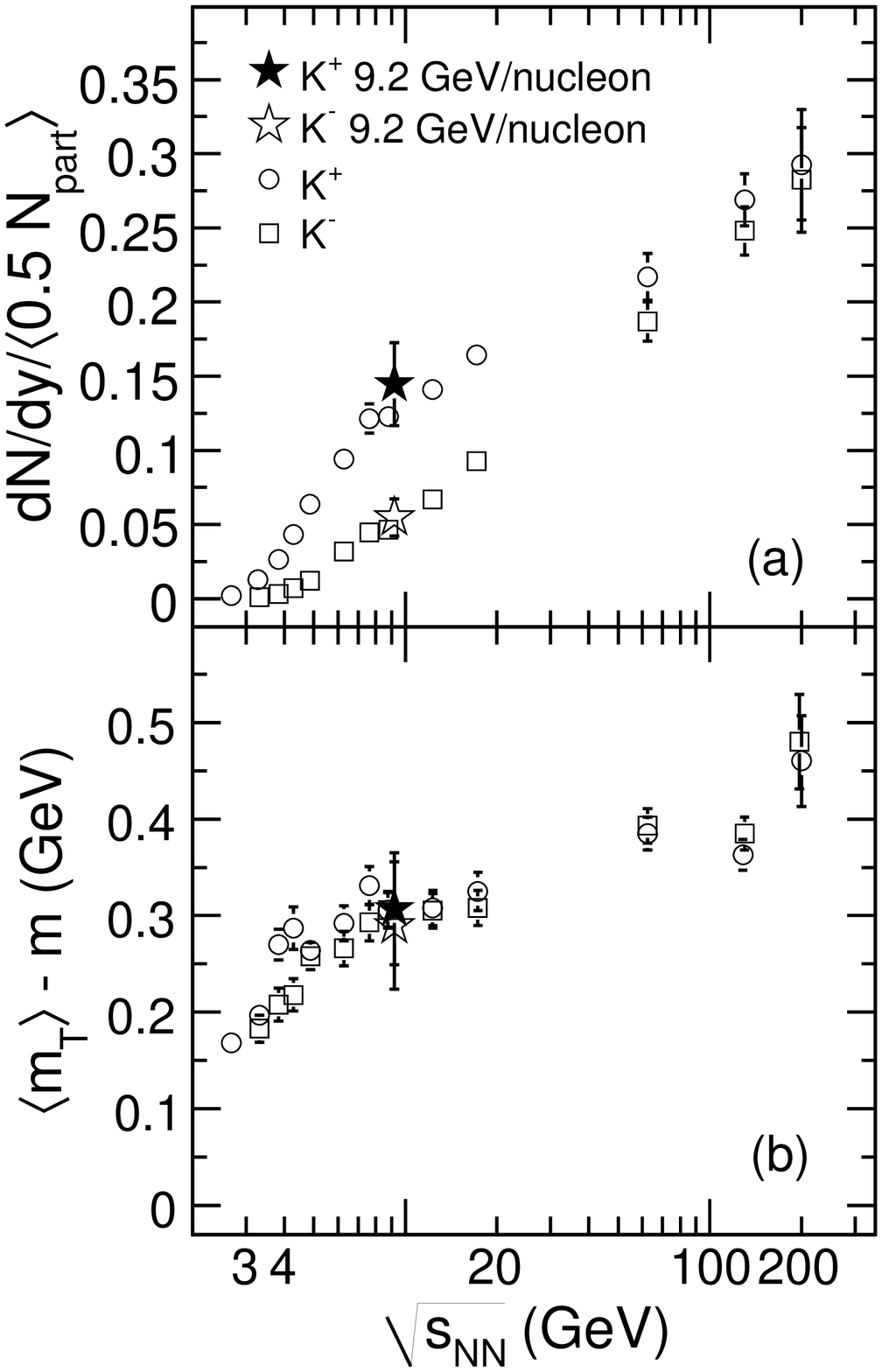}
\caption{(a) $dN/dy$ normalized by 
$\langle N_{\mathrm {part}} \rangle /2$ (b)
$\langle m_{T} \rangle - m$ of $K^{\pm}$, in 0--10\% central 
Au+Au collisions for $\sqrt{s_{NN}} = $ 9.2 GeV compared to previous 
results from AGS~\cite{ags}, SPS~\cite{sps}, and RHIC~\cite{STARPID}. 
The errors shown are the quadrature sum of statistical and systematic
uncertainties.
}
\label{energydndymt_k}
\eef
\subsection{Energy dependence of particle production}
Figure~\ref{dndetaen} shows the $dN_{\mathrm {ch}}/d\eta$ at midrapidity normalized by
$\langle N_{\mathrm {part}} \rangle$/2 as a function of $\sqrt{s_{NN}}$.
The result from $\sqrt{s_{NN}}$ = 9.2 GeV is in agreement with the general energy dependence
trend observed at the AGS~\cite{ags}, SPS~\cite{sps}, and RHIC~\cite{STARPID,PHENIXnch}. 
The result at 9.2 GeV  has a value close to that obtained at a 
similar energy ($\sqrt{s_{NN}}$ = 8.8 GeV)
by the NA49 experiment at SPS~\cite{sps}.
Figures~\ref{energydndymt_pi}(a) and~\ref{energydndymt_k}(a) show $dN/dy$ normalized 
by $\langle N_{\mathrm {part}} \rangle$/2 for $\pi^{\pm}$ and $K^{\pm}$, respectively,
in 0--10\% central Au+Au collisions at $\sqrt{s_{NN}}$ = 9.2 GeV, compared
to previous results at AGS~\cite{ags}, SPS~\cite{sps}, and RHIC~\cite{STARPID}. Within errors,
the yields are consistent
with previous results at similar $\sqrt{s_{NN}}$.
Figures~\ref{energydndymt_pi}(b) and~\ref{energydndymt_k}(b) show the  
$\langle m_{T} \rangle - m$ for $\pi^{\pm}$ and $K^{\pm}$, respectively, in 
0--10\% central Au+Au collisions at $\sqrt{s_{NN}} = $ 9.2 GeV. The results are 
also compared to previous measurements at various energies.
The results from  Au+Au collisions at $\sqrt{s_{NN}}$ = 9.2 GeV are 
consistent with corresponding measurements at SPS energies at 
similar $\sqrt{s_{NN}}$. Both $dN/dy$ and $\langle m_{T} \rangle - m$
are obtained using data in the measured $p_{T}$ ranges and extrapolations assuming
certain functional forms for the unmeasured $p_{T}$ ranges, as discussed in section V.B 
of our previous paper~\cite{STARPID}. For the present midrapidity measurements, the percentage 
contribution to the yields from extrapolation are about 20\% for $\pi^{\pm}$, 
50\% for $K^{\pm}$, and 25\% for $p$. 

The $\langle m_{T}\rangle - m$
values increase with $\sqrt{s_{NN}}$ at lower AGS energies, stay
independent of $\sqrt{s_{NN}}$ at the SPS and RHIC 9.2 GeV collisions, 
then tend to rise further with 
increasing $\sqrt{s_{NN}}$ at the higher beam energies at RHIC. For a thermodynamic 
system, $\langle m_{T}\rangle - m$ can be an approximate representation of
the temperature of the system, and  $dN/dy$ $\propto$ $\ln(\sqrt{s_{NN}})$ 
may represent its 
entropy. In such a scenario, the observations could reflect the characteristic 
signature of a first order 
phase transition, as proposed by Van Hove~\cite{vanhove}. Then the constant value 
of $\langle m_{T}\rangle - m$ vs. $\sqrt{s_{NN}}$ around 9.2 GeV 
has one possible interpretation  in 
terms of 
formation of a mixed phase of a QGP and hadrons during the evolution of the heavy-ion
system. However, there could be several other effects to which $\langle m_{T}\rangle - m$ is 
sensitive,
which also need to be understood for proper interpretation of the data~\cite{bedanga}. 
The energy dependencies of the proton
$dN/dy$ and $\langle m_{T} \rangle - m$ are not discussed in this paper, as the STAR results
are presented without correction for feed down contributions. 
The low event statistics in the present data does not allow us to obtain
feed-down corrections from the data itself.
All results presented in this paper are from inclusive protons and anti-protons
as in our previous paper at higher energies at RHIC~\cite{STARPID}. 

\bef
\includegraphics[scale=0.45]{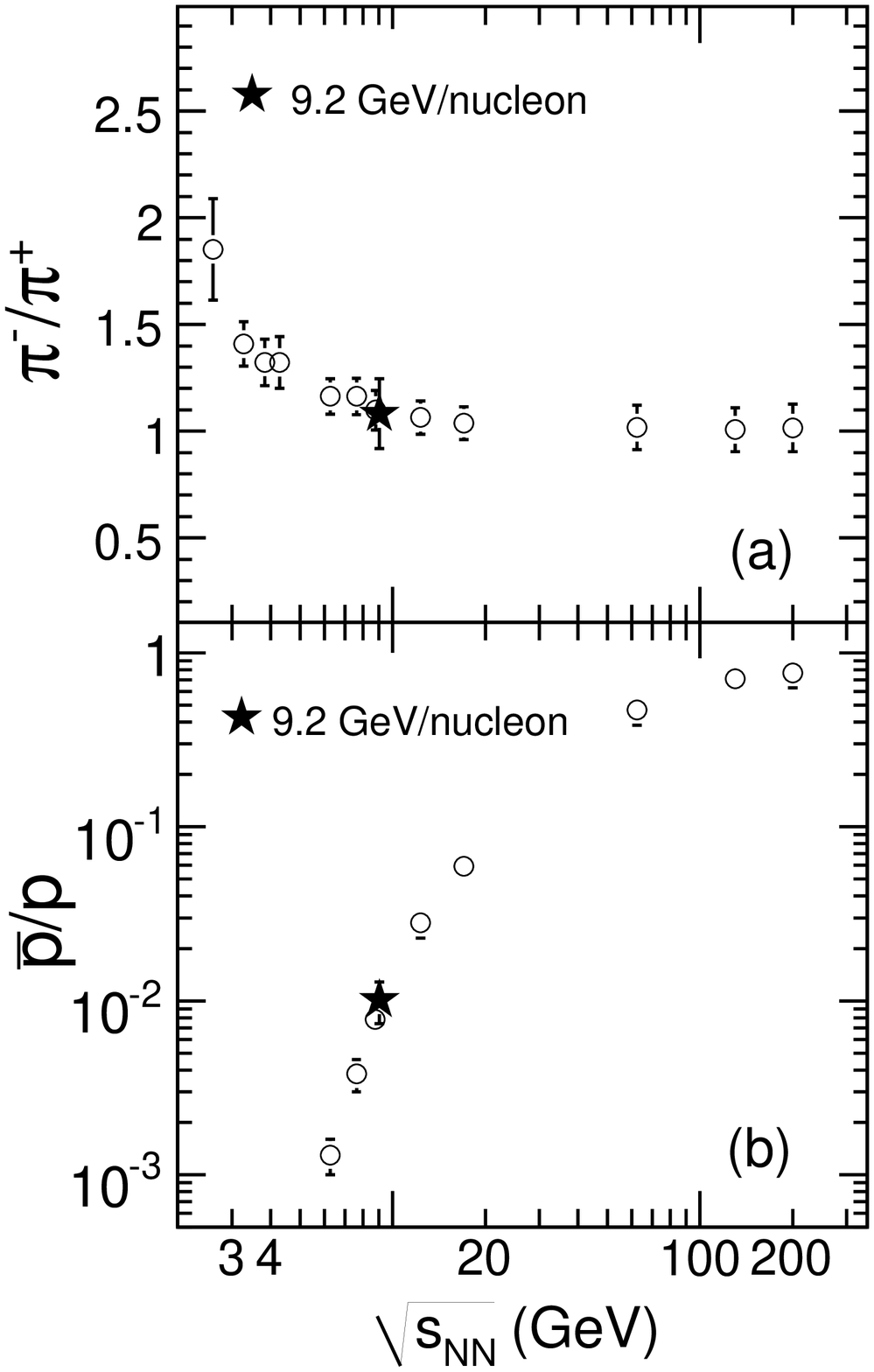}
\caption{(a) $\pi^{-}/\pi^{+}$ and (b) $\bar{p}/p$ ratios at midrapidity 
($\mid y \mid < 0.5$) for central 0--10\% Au+Au collisions at 
$\sqrt{s_{NN}}$ = 9.2 GeV compared to previous results from AGS~\cite{ags}, SPS~\cite{sps}, and 
RHIC~\cite{STARPID}.
The errors shown are the quadrature sum of statistical and systematic
uncertainties.
The systematic errors on $\pi^{-}/\pi^{+}$ and $\bar{p}/p$ for $\sqrt{s_{NN}}$ = 9.2 GeV data 
are 15\% and 27\%, respectively.
}
\label{ratioen_pip}
\eef

\bef
\includegraphics[scale=0.45]{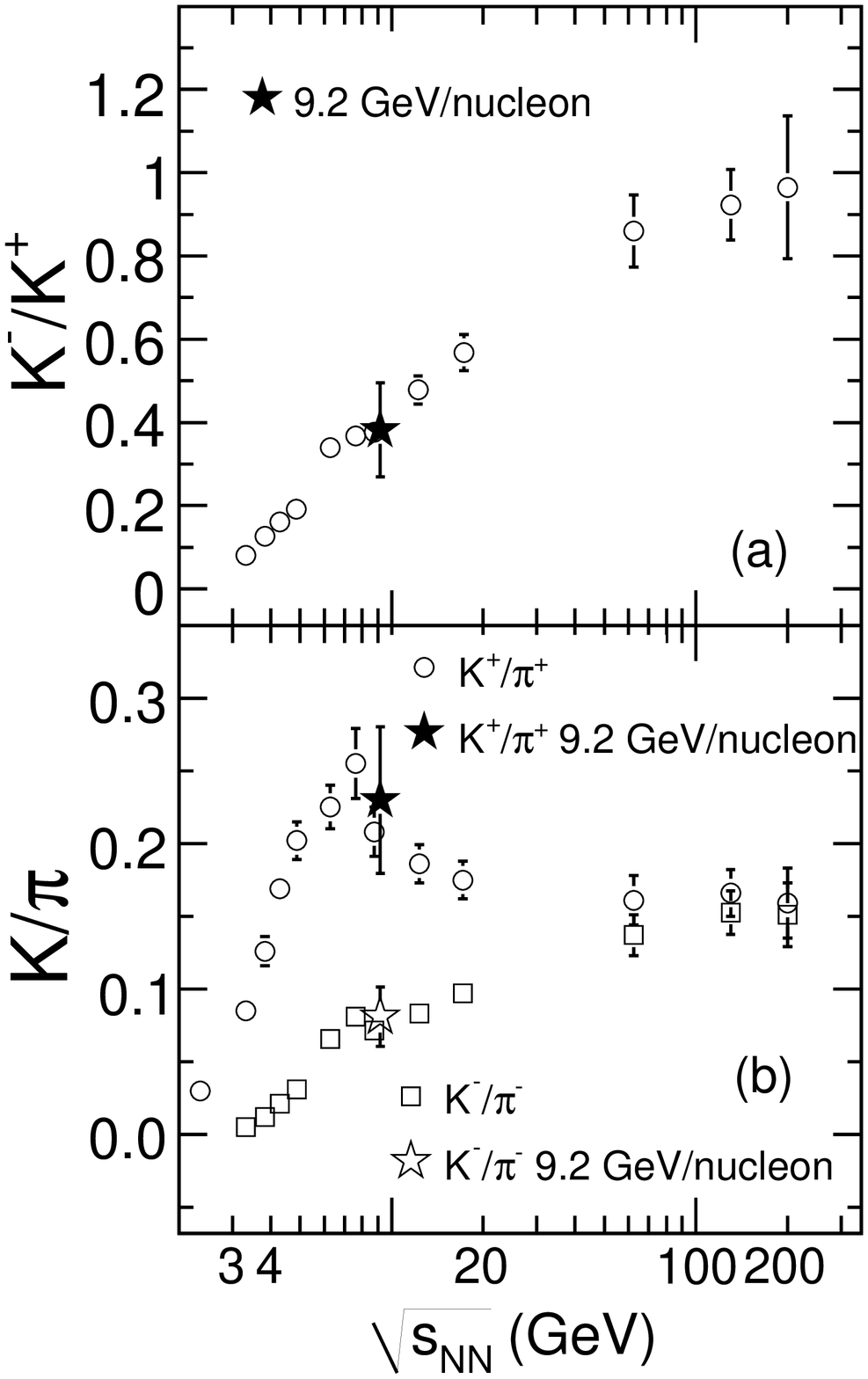}
\caption{(a) $K^{-}/K^{+}$ and (b) $K/\pi$ ratios at midrapidity 
($\mid y \mid < 0.5$) for central 0--10\% Au+Au collisions at 
$\sqrt{s_{NN}}$ = 9.2 GeV compared to previous results from AGS~\cite{ags}, SPS~\cite{sps}, and 
RHIC~\cite{STARPID}.
The errors shown are the quadrature sum of statistical and systematic
uncertainties.
The systematic errors on $K^{-}/K^{+}$ and $K/\pi$ for $\sqrt{s_{NN}}$ = 9.2 GeV data are 23\% and
19\%, respectively.
}
\label{ratioen_k}
\eef

Figures~\ref{ratioen_pip}(a) and~\ref{ratioen_pip}(b) show the collision energy dependence of the 
particle ratios 
$\pi^{-}/\pi^{+}$ and $\bar{p}/p$, respectively, in central heavy-ion collisions. Similarly, 
Figs.~\ref{ratioen_k}(a) and~\ref{ratioen_k}(b) 
show the ratios of $K^{-}/K^{+}$ and $K/\pi$, respectively.
The new results from Au+Au collisions at $\sqrt{s_{NN}}$ = 9.2 GeV
follow the $\sqrt{s_{NN}}$ trend established by previous measurements. 
The $p_{T}$-integrated $\pi^{-}/\pi^{+}$ ratio
at $\sqrt{s_{NN}}$ = 9.2 GeV is 1.08 $\pm$ 0.04 (stat.) $\pm$ 0.16 (sys.). 
Those at lower beam energies have values 
much larger than unity, which could be due to significant contributions from resonance 
decays (such as from $\Delta$ baryons). 
The value of the $\bar{p}/p$ ratio at  $\sqrt{s_{NN}}$ = 9.2 GeV is 
 0.010 $\pm$ 0.001 (stat.) $\pm$ 0.003 (sys.)
indicating large values of net-protons.
The $\bar{p}/p$ ratio increases with
increasing collision energy and approaches unity for top RHIC energies. This indicates that
at higher beam energies the $p$ ($\bar{p}$)
production at midrapidity is dominated by pair production. 
The $K^{-}/K^{+}$ ratio at $\sqrt{s_{NN}} =$ 9.2 GeV is 
0.38 $\pm$ 0.05 (stat.) $\pm$ 0.09 (sys.), 
indicating a significant contribution to kaon production from associated production 
at lower collision energies. With increasing $\sqrt{s_{NN}}$, the $K^{-}/K^{+}$ ratio
approaches 
unity, indicating dominance of
kaon pair production.
The $K/\pi$ ratio is of interest, as it expresses the enhancement of strangeness production 
relative to non-strange hadrons in heavy-ion collisions compared to $p+p$ collisions.
The increase in $K^{+}/\pi^{+}$ ratio with beam energies up to 
$\sqrt{s_{NN}} =$ 7.7 GeV at SPS and the subsequent decrease and possible 
saturation 
with increasing beam energies has been a subject of intense theoretical debate 
recently~\cite{sps, kpitheory}. 
The discussions 
mainly focus on the question of the relevant degrees of freedom that are necessary 
to explain the energy dependence of the $K/\pi$ ratio. Our new results from Au+Au collisions at 
$\sqrt{s_{NN}}$ = 9.2 GeV with only about 3000 events (hence with large errors) are found to be 
consistent 
with the previously observed energy dependence. 

\subsection{Azimuthal anisotropy}
The study of collective flow in relativistic nuclear
collisions could provide insights into the equation of state (EOS) of the
matter created by the collisions. As discussed earlier, there are two types of
azimuthal anisotropy that are widely studied in heavy-ion collisions, 
directed flow ($v_{1}$) and elliptic flow ($v_{2}$). 
Directed flow measurements at forward rapidities describe
the ``side-splash'' motion of the collision products.
Hence, it is an
important tool to probe the dynamics of the system at forward rapidities~\cite{collective}.
Since $v_{1}$ is generated very early in the evolution 
of heavy-ion collisions, it probes the onset of bulk 
collective dynamics. 
The shape of 
$v_1$ vs. rapidity around midrapidity 
is suggested as a signature of a first order
phase transition~\cite{directed}. On the other hand, the characterization of the 
elliptic flow of produced particles by their azimuthal anisotropy has proven to be one 
of the most successful probes of the dynamics in  Au+Au
collisions at RHIC~\cite{flow1,flow2,flow3,flow4,flow5,flow6,flow7,flow8}.
Elliptic flow provides the possibility to gain information about the degree of thermalization
of the hot, dense medium. 
Studying its dependence on system size,
number of constituent quarks, transverse momentum, and transverse mass, is crucial to the 
understanding of the properties of the produced matter.

\bef
\begin{center}
\includegraphics[scale=0.39]{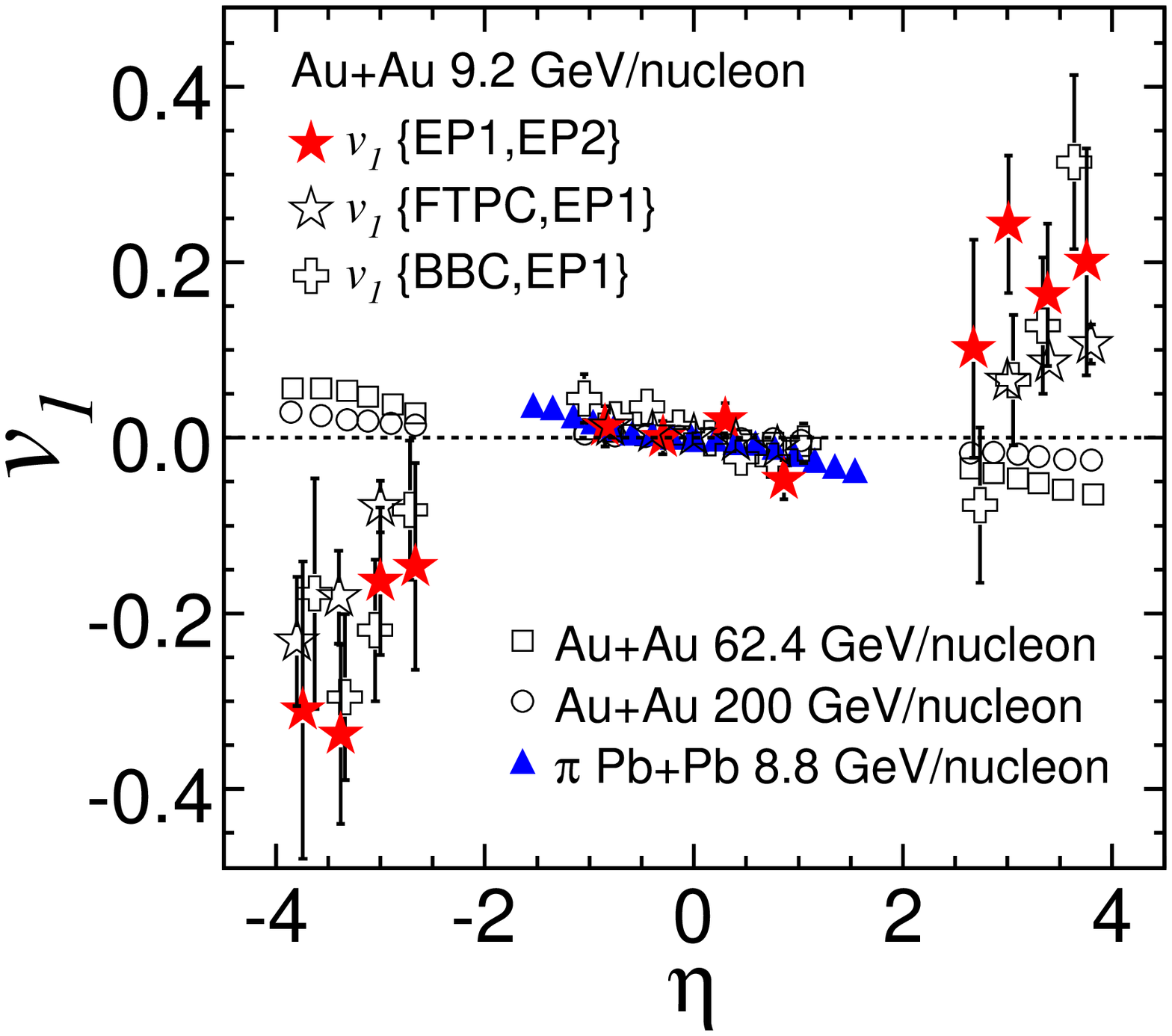}
\caption{(Color online) Charged hadron $v_{1}$ vs. $\eta$
from the 0--60\% collision centrality Au+Au collisions at $\sqrt{s_{NN}}$ = 9.2 GeV.
The errors shown are statistical.
Systematic errors are discussed in section II H. 
The solid star symbols are the results obtained from the mixed harmonic method,
while the open star and open plus symbols represent results from the standard methods (see
text for details).
The results are compared to 
$v_{1}$ from 30--60\% collision centrality Au+Au collisions at $\sqrt{s_{NN}} = $ 62.4 and 
200 GeV~\cite{v14systempaper}. 
For comparison, $v_{1}$ for charged pions 
for the 0--60\% collision centrality from Pb+Pb collisions at $\sqrt{s_{NN}}$ = 8.8 GeV
are also shown~\cite{v2NA49}.
}
\label{figv11}
\end{center}
\eef

\bef
\begin{center}
 \includegraphics[scale=0.39]{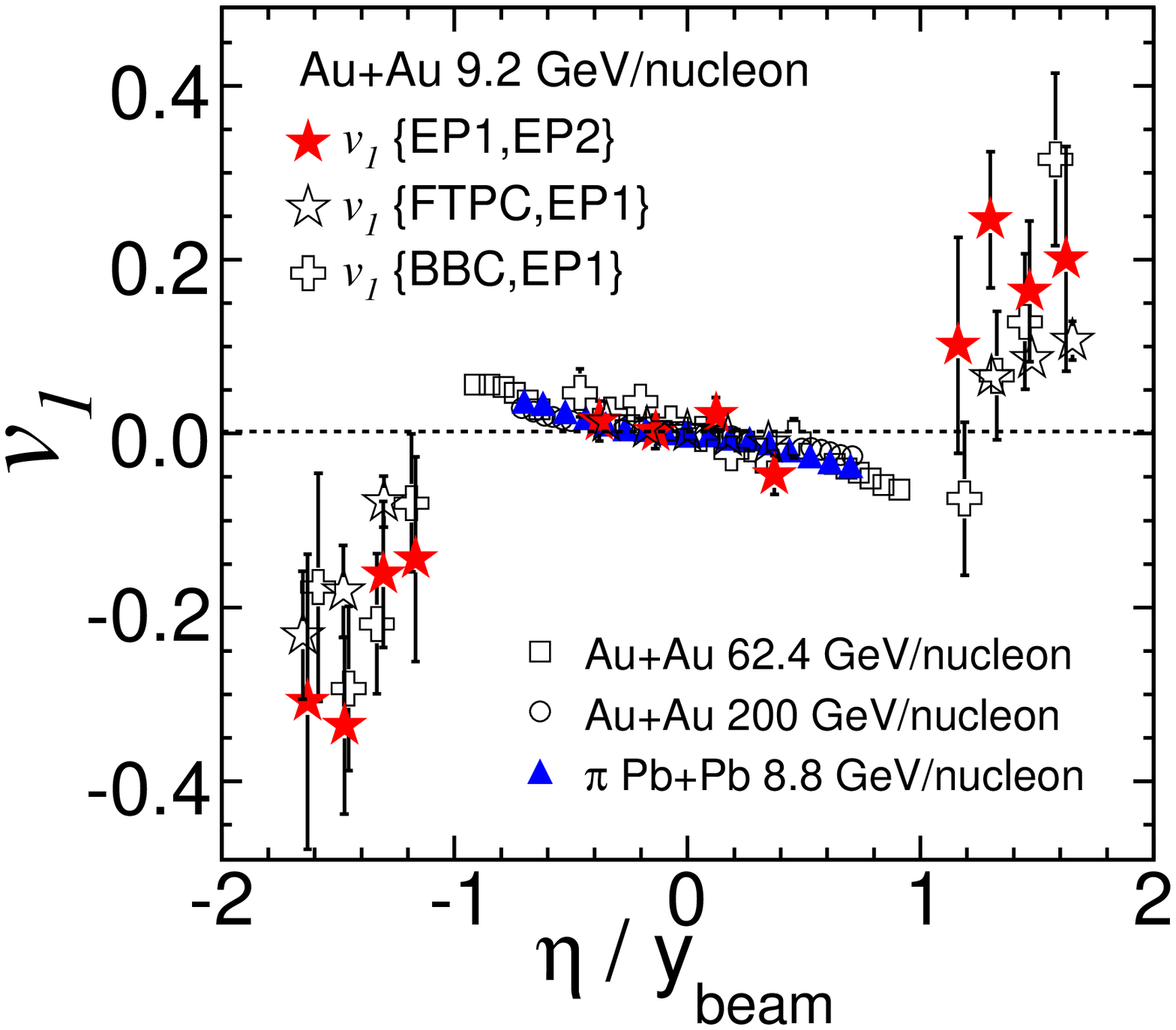}
\caption{(Color online) Same as Fig.~\ref{figv11}, but plotted as a function of $\eta/y_{\rm{beam}}$.
}
\label{figv12}
\end{center}
\eef

Figure~\ref{figv11} shows charged hadron $v_{1}$ results in Au+Au collisions
for the 0--60\% collision centrality at $\sqrt{s_{NN}}$~=~9.2 GeV, compared to
corresponding results from 30--60\% central Au+Au collisions at $\sqrt{s_{NN}}$ = 62.4 and
200 GeV~\cite{v14systempaper}. The $p_{T}$ range of this study is 
0.15--2.0 GeV/$c$.
The $v_{1}$ results from Au+Au collisions at $\sqrt{s_{NN}}$~=~9.2 GeV are shown 
for the three different methods, as described in section II E. The results
from the three methods are consistent within the
error bars.
These results are also compared with $v_{1}$ for charged 
pions in Pb+Pb collisions at $\sqrt{s_{NN}}$ = 8.8 GeV measured by NA49~\cite{v2NA49}. 
At midrapidity, all the results have comparable values.
At forward rapidity ($|\eta| >$ 2), the trend of $v_{1}$ for higher $\sqrt{s_{NN}}$
(62.4 and 200 GeV) appears to be different from that for $\sqrt{s_{NN}}$ = 9.2 GeV.
This can be explained by contributions from spectator protons 
to the directed flow signal at large $|\eta|$. The beam rapidities ($y_{\rm{beam}}$)
for $\sqrt{s_{NN}}$ = 9.2, 62.4, and 200 GeV are 2.3, 4.2, and 5.4 respectively.
With $\eta$ divided by the respective $y_{\rm{beam}}$ values 
for the beam energies
(Fig.~\ref{figv12}), all the $v_{1}$ values follow a common trend for the 
measured $|\eta|/y_{\rm{beam}} < 1$ range.

\bef
\begin{center}
\includegraphics[scale=0.39]{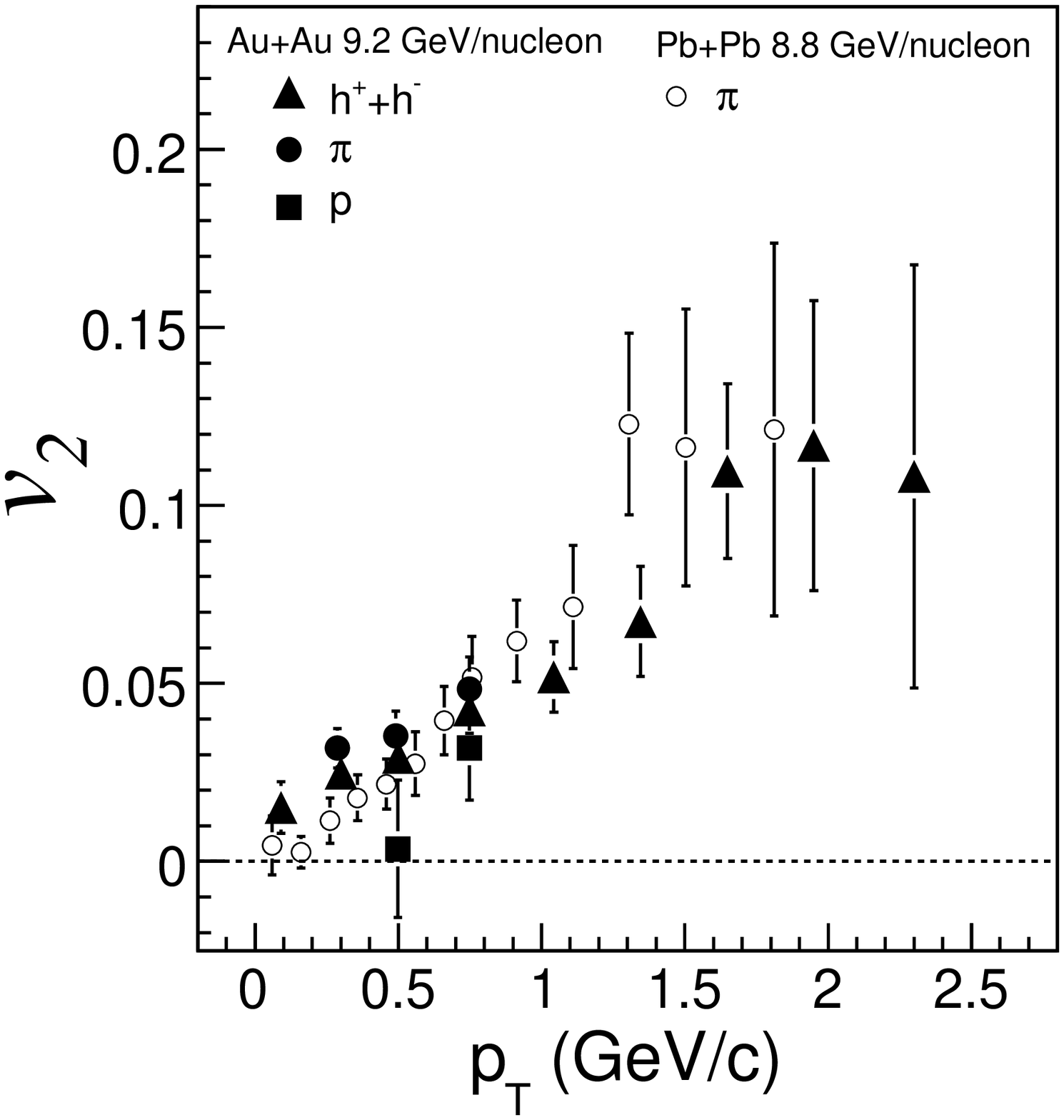}
\caption{$v_{2}$ as a function of $p_T$ for charged
hadrons (solid triangles), $\pi$ (solid circles), and $p$
(solid squares) in 0--60\% Au+Au collisions at 
$\sqrt{s_{NN}} =$ 9.2 GeV. 
The error bars include only statistical uncertainties for 
$\sqrt{s_{NN}} =$ 9.2 GeV data.
The corresponding systematic error is discussed in section II H.
For comparison, $v_{2}$($p_{T}$) results for $\pi$ (open circles) from
NA49~\cite{v2NA49} in 0--43.5\% Pb + Pb collisions at 
$\sqrt{s_{NN}} =$ 8.8 GeV, are also shown.
}
\label{figv21}
\end{center}
\eef

\bef
\begin{center}
\includegraphics[scale=0.39]{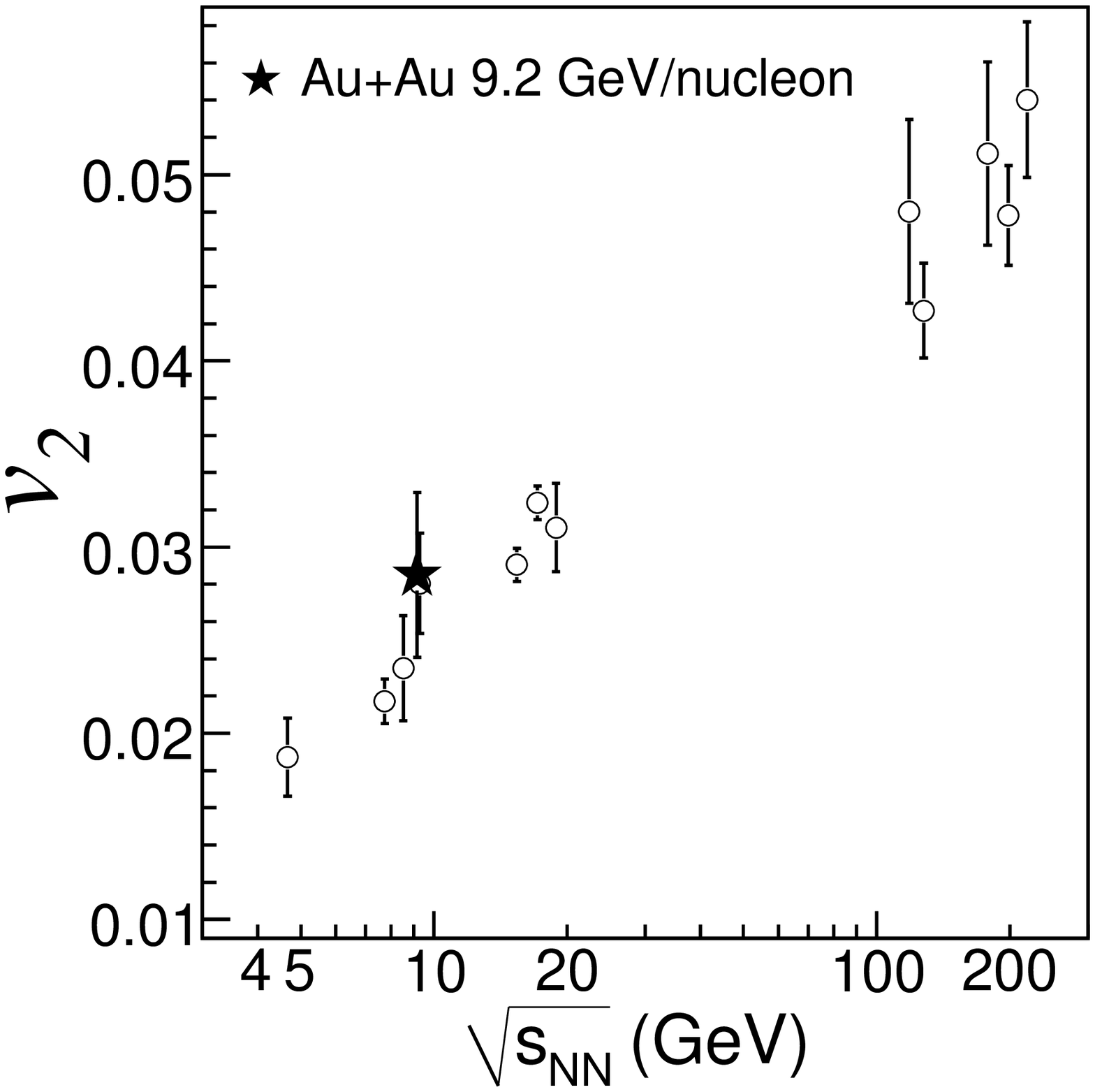}
\caption{Energy dependence of $v_{2}$ near
midrapidity ($-1 < \eta < 1$) for $\sqrt{s_{NN}} =$ 9.2 GeV 
0--60\% central Au+Au collisions.
Only statistical errors are shown. The results of STAR charged hadron
$v_2$~\cite{STAR} are compared to those measured by
E877~\cite{E877}, NA49~\cite{v2NA49}, PHENIX~\cite{PHENIX}, and 
PHOBOS~\cite{flow2, flow6, PH0BOS}.
}
\label{figv22}
\end{center}
\eef
Figure~\ref{figv21} shows $v_{2}$($p_{T}$) for charged hadrons, pions, and
protons in $\sqrt{s_{NN}} = $ 9.2 GeV 
collisions. For comparison, we show pion $v_{2}$ results from NA49~\cite{v2NA49} 
at similar $\sqrt{s_{NN}}$. Within the
statistical errors, there is good agreement between results from the two experiments.
At top RHIC energies, $v_2$ at low $p_T$ shows a characteristic scaling with particle 
mass~\cite{starv2} that is consistent with hydrodynamic behavior; however, the available 
statistics in the current analysis are insufficient to extend this study to 9.2 GeV.  
The small number of events also precludes the extension
of the measurements to larger $p_{T}$ values, to
study the number of constituent quark scaling of $v_{2}$ observed at  $\sqrt{s_{NN}}$ = 200 GeV.
Figure~\ref{figv22} shows the elliptic flow parameter at $\sqrt{s_{NN}} = $ 9.2 GeV 
compared to other beam energies~\cite{E877,v2NA49,PHENIX,flow2, flow6, PH0BOS}.
The STAR data at $\sqrt{s_{NN}} =$ 9.2 GeV, denoted by the
star symbol, follow the observed $\sqrt{s_{NN}}$ dependence.

\subsection{Pion interferometry}
\bef
\includegraphics[scale=0.4]{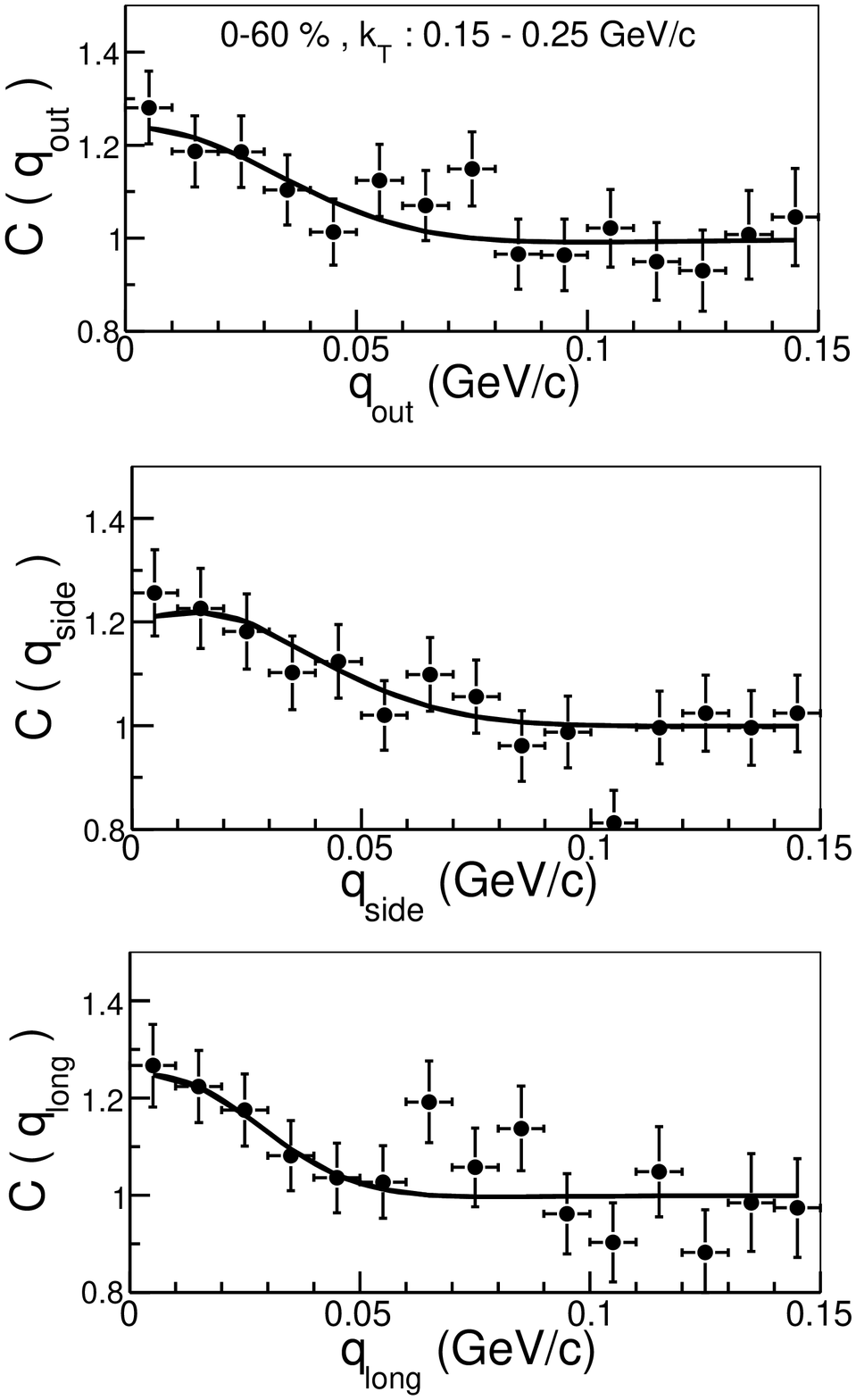}
\caption{Projections of the three-dimensional correlation function and
corresponding Bowler-Sinyukov~\cite{lednicky} fits (lines) for negative pions 
from the 0--60$\%$ central Au+Au events 
and $k_{T}$ $=$ [150, 250] MeV/$c$.
}
\label{hbt}
\eef

Information about the space-time structure of the emitting source
can be extracted with intensity interferometry techniques~\cite{Goldhaber:1960sf}. 
The primary goal of pion interferometry, performed at midrapidity and 
at low transverse momentum, is to study the space-time size of the emitting 
source and freeze-out processes of the dynamically evolving collision fireball. 
The 3-dimensional correlation functions are fitted with Eq.~(\ref{eq:two}),
where $R_{i}$ is the homogeneity length in the $i$ 
direction~\cite{Bertsch:1988db,Pratt:1986cc,Chapman:1994yv}.
Projections of the fit to the correlation function of the 0--60\% most central collisions, 
weighted
according to the mixed-pair background, are shown in Fig.~\ref{hbt}.
The three panels show the projections of the 3-dimensional 
correlation function onto the $q_{\rm{out}}$, $q_{\rm{side}}$, and $q_{\rm{long}}$ axes.
The curves show Bowler-Sinyukov fits~\cite{lednicky} to the Coulomb-corrected correlation 
function.
Table~\ref{table4} lists the HBT parameters obtained from the fits along with statistical 
errors.

\begin{table}
\caption{\label{table4}The HBT parameters for 0--60$\%$ central events and 
$150 < k_{T} < 250$ MeV/$c$.}
\vspace{0.15cm}
\begin{tabular}{c|c|c|c}
\hline
$\lambda$&$R_{\rm out}$ (fm) &$R_{\rm side}$ (fm) &$R_{\rm long}$ (fm) \\[0.3mm]
\hline
0.36 $\pm$ 0.08  &  5.05 $\pm$ 0.96 & 3.52 $\pm$ 0.56  &  3.25 $\pm$ 0.86  \\ [0.2mm]
\hline
\end{tabular}
\end{table}

The radius parameter $R_{\rm side}$ has the most direct correlation with
the source geometry, whereas $R_{\rm out}$ encodes both geometry
and time scale. Hydrodynamic calculations with a first order phase transition predict a
ratio of $R_{\rm out}$/$R_{\rm side}$  larger than unity. Our measurements indicate the ratio 
$R_{\rm out}$/$R_{\rm side}$ = 1.4 $\pm$ 0.4. 

\section{Freeze-out parameters and Phase diagram}

The measured hadron spectra reflect the properties of the bulk matter at kinetic freeze-out, 
after
elastic collisions among the hadrons have ceased. 
More direct information on the earlier
stages can be deduced from the integrated yields of the different hadron species, which 
change only via
inelastic collisions. The point in time at which these inelastic collisions cease is referred to as 
chemical freeze-out,
which takes place before kinetic freeze-out. The transverse momentum distributions of the
different particles contain two components, one random and one collective.
The random component can be
identified as the one that depends on the temperature
of the system at kinetic freeze-out ($T_{\mathrm {kin}}$). The collective component, 
which arises
from the matter density gradient from the center to the boundary of the fireball created in 
high energy
nuclear collisions, 
is generated by
collective flow in the transverse direction, and 
is characterized by its velocity $\beta_T$.

\bef
\includegraphics[scale=0.39]{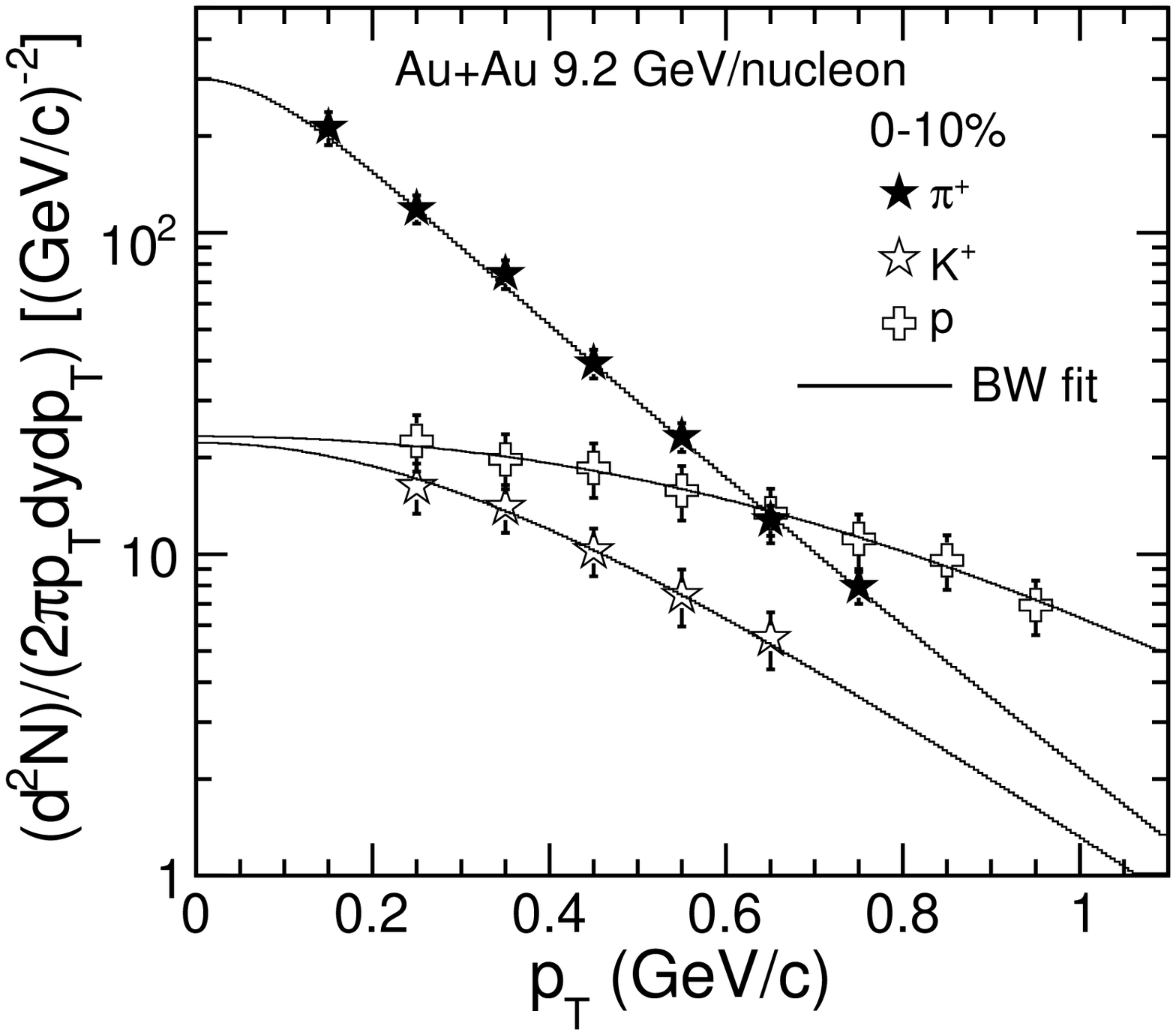}
\caption{Midrapidity transverse momentum distributions of pions, kaons, and protons
(no feed-down correction) for 0--10\% 
most central Au+Au collisions at $\sqrt{s_{NN}}$ = 9.2 GeV, fitted to blast-wave model 
calculations~\cite{heniz}.
The extracted kinetic freeze-out parameters are  
$T_{\mathrm {kin}}$ = 105 $\pm$ 10 (stat.) $\pm$ 16 (sys.) MeV and 
$\langle \beta_T \rangle$ = 0.46$c$ $\pm$ 0.01$c$ (stat.) $\pm$ 0.04$c$ (sys.).
}
\label{bw}
\eef
Assuming that the system attains thermal equilibrium, the blast-wave (BW)
formulation~\cite{heniz} can be used to extract $T_{\mathrm {kin}}$ and $\langle \beta_T \rangle$. 
The transverse flow velocity of a particle at a distance $r$ from the center of the emission 
source, as a function of the surface velocity ($\beta_s$) of the expanding cylinder, 
is parameterized as
$\beta_T(r) = \beta_s (r/R)^n$, where $n$ is found by fitting the data. 
The transverse momentum spectrum is then
\begin{eqnarray}
\frac{dN}{p_T \, dp_T} & \propto & \int_0^R r \, dr \, m_T
I_0\left(\frac{p_T \sinh\rho(r)}{T_{\rm{kin}}}\right) \nonumber \\
& & \times K_1\left(\frac{m_T \cosh\rho(r)}{T_{\rm{kin}}}\right),
\label{blasteq}
\end{eqnarray}
where $I_0$ and $K_1$ are modified Bessel functions and $\rho(r) = 
\tanh^{-1}{\beta_T(r)}$. Simultaneous fits to the $p_{T}$ distributions of $\pi$, $K$, 
and $p$
at midrapidity for central 0--10\% Au+Au collisions at  $\sqrt{s_{NN}}$ = 9.2 GeV 
are shown in 
Fig.~\ref{bw}. 
The extracted  parameters are 
$T_{\mathrm {kin}}$ = 105 $\pm$ 10 (stat.) $\pm$ 16 (sys.) MeV,
$\langle \beta_T \rangle$ = 0.46$c$ $\pm$ 0.01$c$ (stat.) $\pm$ 0.04$c$ (sys.),
and $n$ = 0.9 $\pm$ 6.4 (stat.) $\pm$ 6.4 (sys.)
with $\chi^{2}/\rm{ndf}$ $=$ 15/17. 
The parameter $n$ is poorly constrained by the fits in this low event statistical data set.
The parameter
values do not change within the quoted errors for other centrality ranges. 
Only statistical errors
are used for obtaining the fit parameters shown in the figure. 
Inclusion of  systematic errors gives similar values of $T_{\mathrm {kin}}$  and 
$\langle \beta_{T} \rangle$.
Similar studies have been done
for other higher energy collisions at RHIC~\cite{STARPID}.

\bef
\includegraphics[scale=0.39]{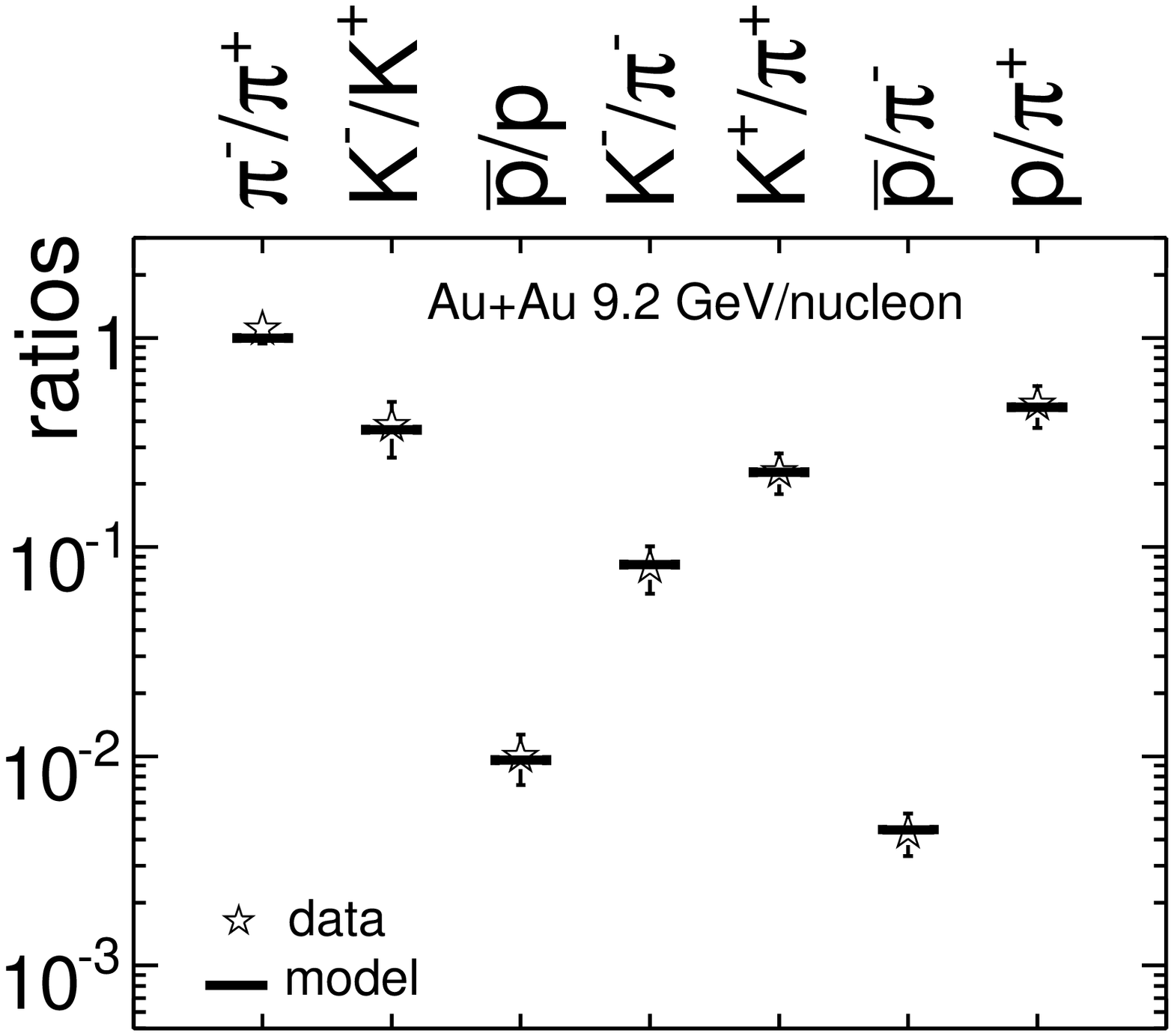}
\vspace{-1cm}
\caption{Midrapidity particle ratios for 0--10\% most central Au+Au
collisions at $\sqrt{s_{NN}}$ = 9.2 GeV, fitted to thermal model calculations. See text
for details. 
The extracted chemical freeze-out temperature  is 
$T_{\mathrm {ch}}$ = 151 $\pm$ 2 (stat.) $\pm$ 7 (sys.) MeV 
and baryon chemical potential is $\mu_{\mathrm B}$ = 354 $\pm$ 7 (stat.) $\pm$ 30 (sys.) MeV.
}
\label{chem}
\eef

Within a statistical model in thermodynamical equilibrium, the particle abundance in a 
system of volume $V$ can be given by
\begin{equation}
N_i/V=\frac{g_i}{(2\pi)^3}\gamma_{S}^{S_i}\int\frac{1}{\exp\left(\frac{E_i-\mu_BB_i-\mu_SS_i}{T_{\rm {ch}}}\
\right)\pm 1}d^3p\,,
\label{eq:chemical}
\end{equation}
where $N_i$ is the abundance of particle species $i$, $g_i$ is the spin degeneracy, $B_i$ and $S_i$
are the baryon number and strangeness number, respectively, $E_i$ is the particle energy, and the 
integral is taken over all momentum space~\cite{STARPID}. The model parameters are the chemical 
freeze-out 
temperature ($T_{\mathrm {ch}}$), 
the baryon ($\mu_{\mathrm {B}}$) and strangeness ($\mu_{S}$) chemical potentials, 
and the
{\it ad hoc}
strangeness suppression factor ($\gamma_{S}$).
Measured particle ratios are used to constrain the values of temperature 
($T_{\mathrm {ch}}$) and baryon chemical potential ($\mu_{\mathrm B}$) at chemical freeze-out using 
the statistical model assumption that the system is in thermal and chemical equilibrium at that 
stage. 
Fits are performed to the various ratios for midrapidity central 0--10\% Au+Au collisions at 
$\sqrt{s_{NN}}$ = 9.2 GeV using such a model, and are shown in Fig.~\ref{chem}.
The analysis is done within the framework of a statistical model as discussed in Ref.~\cite{peter}. 
This
model has been used to extract chemical freeze-out parameters at higher RHIC energies~\cite{STARPID}.
The extracted parameter values are 
$T_{\mathrm {ch}}$ = 151 $\pm$ 2 (stat.) $\pm$ 7 (sys.) MeV,
$\mu_{\mathrm B}$ = 354 $\pm$ 7 (stat.) $\pm$ 30 (sys.) MeV,
$\mu_{S}$ = 25 $\pm$ 9 (stat.) $\pm$ 14 (sys.) MeV, 
and $\gamma_{S}$ = 0.9 $\pm$ 0.7 (stat.) $\pm$ 0.1 (sys.) 
for 9.2 
GeV data. These values are
very close to those extracted from the measurements at SPS for similar 
$\sqrt{s_{NN}}$~\cite{spsfit}.
Only  statistical errors on the particle production ratios are used for obtaining the fit 
parameters. Inclusion of systematic
errors gives similar values of $T_{\mathrm {ch}}$ and $\mu_{\mathrm {B}}$.
\bef
\includegraphics[scale=0.47]{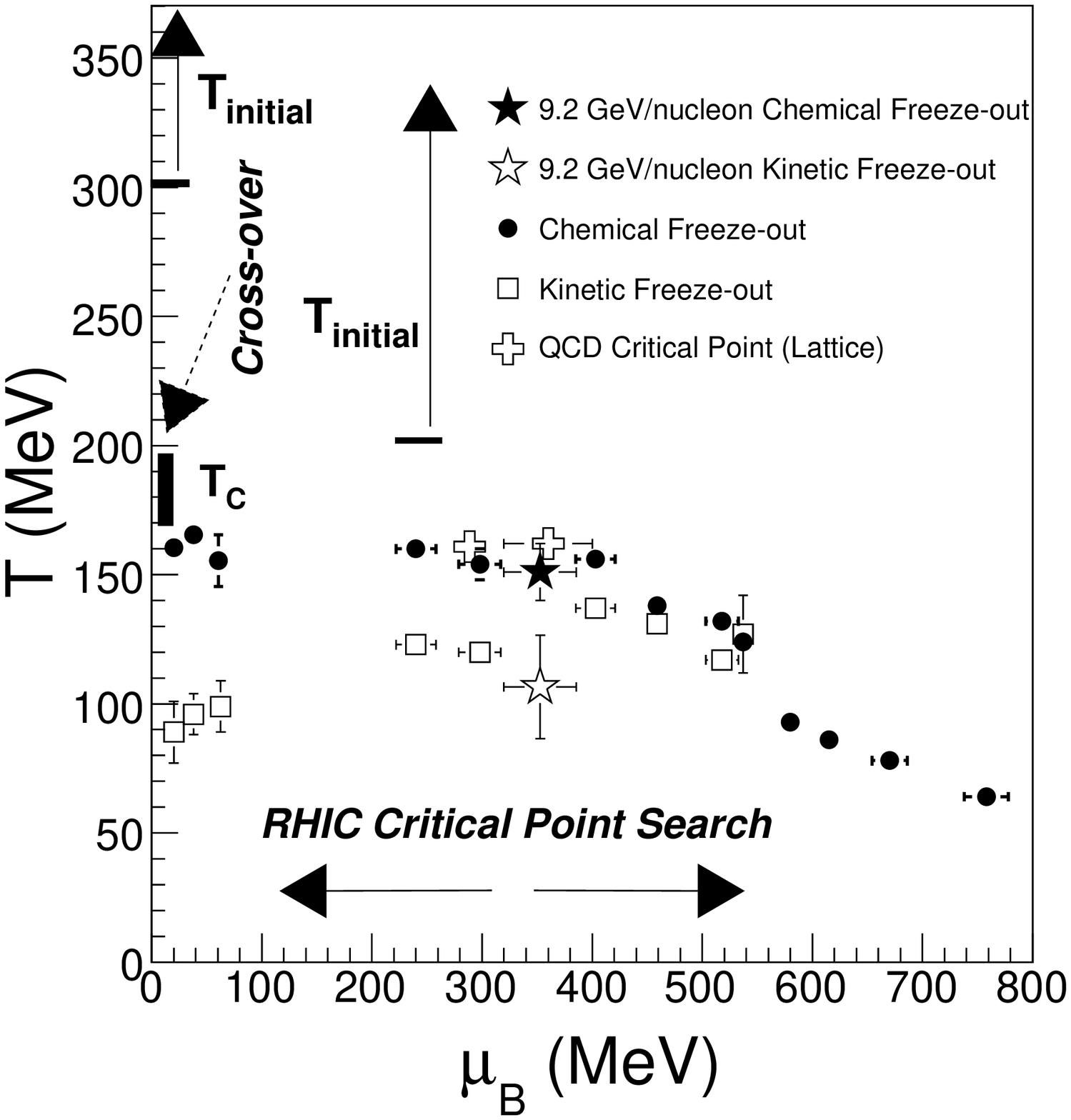}
\caption{Temperature vs. baryon chemical potential ($\mu_{\mathrm {B}}$) from heavy-ion
collisions at various $\sqrt{s_{NN}}$~\cite{STARPID}. 
The $\mu_{\mathrm {B}}$ values shown are estimated at chemical freeze-out.
The kinetic and chemical freeze-out
parameters, extracted using models assuming thermal and chemical equilibrium 
from midrapidity measurement in central 0--10\% Au+Au collisions at $\sqrt{s_{NN}}$ = 9.2 GeV, 
are shown as star symbols. 
The range of critical temperatures ($T_{\mathrm {c}}$)~\cite{transition} of the 
cross-over quark-hadron phase transition at $\mu_{\mathrm {B}}$ = 0~\cite{crossover}
and the QCD critical point from two different calculations~\cite{qcdcp}
from lattice QCD are also indicated.
Model-based estimates of the range of initial temperature ($T_{\mathrm {initial}}$) 
achieved in heavy-ion collisions 
based in part on direct photon data at top RHIC~\cite{phenixph} and
SPS~\cite{wa98ph}
energies are also shown. 
The range of $\mu_{\mathrm {B}}$ to be scanned in the upcoming RHIC critical point search 
and Beam Energy Scan 
program corresponding
to $\sqrt{s_{NN}}$ = 5.5 to 39 GeV is indicated by horizontal arrows near the $\mu_{\mathrm {B}}$ axis~\cite{bes}.
}
\label{phase}
\eef

Figure~\ref{phase} shows the temperatures at various stages in heavy-ion collisions
as a function of $\mu_{\mathrm {B}}$ (at different $\sqrt{s_{NN}}$). The $\mu_{\mathrm {B}}$ values
shown are estimated at chemical freeze-out. 
The initial temperatures ($T_{\mathrm {initial}}$) achieved at top RHIC and SPS
energies are obtained from models~\cite{photonmod} that explain the direct photon 
measurements from the PHENIX experiment at RHIC~\cite{phenixph} and from the WA98 experiment
at SPS~\cite{wa98ph}. 
From these models, which assume that thermalization is achieved
in the collisions within a time between 0.1--1.2 fm/$c$, the  
$T_{\mathrm {initial}}$ extracted is greater than 300 MeV at RHIC and 
greater than 200 MeV at SPS.
The $T_{\mathrm {ch}}$ and $T_{\mathrm {kin}}$ values extracted from particle ratios
and $p_{T}$ spectra of various hadrons, respectively, from models assuming thermodynamical
equilibrium are also shown. 
The values for $\sqrt{s_{NN}} = $ 9.2 GeV are 
from the data presented in this paper. The values at other $\sqrt{s_{NN}}$ 
are from Ref.~\cite{STARPID} and references therein. It is interesting to observe 
that $T_{\mathrm {ch}}$ and $T_{\mathrm {kin}}$ values approach each other in the high
$\mu_{\mathrm {B}}$ regime. 
A few recent predictions
from lattice QCD calculations~\cite{bmqm09} are also shown in Fig.~\ref{phase}. Several
lattice QCD calculations indicate that the partonic to hadronic phase transition occurs around
$T_{\mathrm {c}}$ $\sim$ 170--190 MeV~\cite{transition}. These calculations also suggest that
the phase transition at $\mu_{\mathrm {B}}$ = 0 is a cross-over~\cite{crossover}. 
Most QCD-based 
model calculations~\cite{phasedia,firstorder} suggest that the phase transition at 
large $\mu_{\mathrm {B}}$ is of first-order.
Two estimates of the
QCD critical point~\cite{qcdcp} in the $T-\mu_{\mathrm {B}}$ plane taking 
$T_{\mathrm {c}}$ = 176 MeV are shown in Fig.~\ref{phase}.
The region planned to be explored in the critical point search program at RHIC is
shown in Fig.~\ref{phase}.

\section{Summary and Outlook}

We have presented measurements of identified particle production,
azimuthal anisotropy, and pion interferometry in
Au+Au collisions at $\sqrt{s_{NN}}$ = 9.2 GeV. The results are obtained from only about 3000 events 
from the lowest beam energy run to date at the RHIC facility.
The transverse momentum spectra of pions, kaons, and protons are presented
for 0--10\%, 10--30\%, 30--60\%, and 0--60\% collision centrality classes.
The bulk properties are studied by measuring the identified hadron $dN/dy$, 
$\langle p_{T} \rangle$, particle ratios, $v_{1}$ (also at forward rapidity), 
$v_{2}$, and HBT radii ($R_{\rm out}$, $R_{\rm side}$, and $R_{\rm long}$).
All measurements are consistent with corresponding previous results from fixed target
experiments at similar $\sqrt{s_{NN}}$. 

The $\langle p_{T} \rangle$ for protons is higher than that for pions,
indicating some degree of collective flow in the radial direction. However, the difference
between $\langle p_{T} \rangle$ for protons and kaons is considerably smaller
at $\sqrt{s_{NN}} =$ 9.2 GeV than at $\sqrt{s_{NN}}$ = 62.4 and 200 GeV at RHIC.
This suggests that the average collective velocity in the radial direction at the lower beam 
energy is 
smaller compared to 62.4 and 200 GeV collisions. 

The $\bar{p}/p$ ratio at midrapidity 
for $\sqrt{s_{NN}} = $ 9.2 GeV collisions is much smaller, with a value of 
0.010 $\pm$ 0.001 (stat.) $\pm$ 0.003 (sys.),
and the $p/\pi^{+}$ ratio is larger compared to Au+Au collisions 
at $\sqrt{s_{NN}} =$ 200 GeV. These measurements indicate large net-proton
density at midrapidity  
in collisions at $\sqrt{s_{NN}} =$ 9.2 GeV. In this region of high net-baryon
density for 9.2 GeV collisions, the dominant channel for kaon production is 
the associated production. The $K^{-}/K^{+}$ ratio has a value of 
0.38 $\pm$ 0.05 (stat.) $\pm$ 0.09 (sys.)
and  the $K^{+}/\pi^{+}$ ratio is slightly higher compared to that in collisions at 
$\sqrt{s_{NN}} =$ 200 GeV.

The directed flow measurements, plotted as a function of pseudorapidity scaled by the 
beam rapidity, have similar values for three collision energies 
($\sqrt{s_{NN}} =$ 9.2, 62.4, and 200 GeV). A large $v_{1}$ signal is observed at forward 
rapidities at $\sqrt{s_{NN}} =$ 9.2 GeV. These collisions could have significant 
contribution from protons that dominate at large $|\eta|$ (spectator effects). 
The $v_{2}$ measurements for charged hadrons, pions, and protons are also presented for 
$\sqrt{s_{NN}} =$ 9.2 GeV Au+Au collisions at RHIC. 
The charged pion $v_{2}$ as a function of $p_{T}$ is observed to be comparable with that 
from NA49 at similar collision energy. The STAR data at $\sqrt{s_{NN}} =$ 9.2 GeV 
are also found to follow the existing beam energy
dependence of $v_{2}$ for charged hadrons. 

The pion interferometry results 
give information of the size of the homogeneity region of the source. The pion HBT radii 
$R_{\rm out}$, $R_{\rm side}$, and $R_{\rm long}$ 
have values 5.05 $\pm$ 0.96 fm, 3.52 $\pm$ 0.56 fm, and 3.25 $\pm$ 0.86 fm,
respectively.

The kinetic freeze-out parameters are extracted from a blast-wave model fit to pion, kaon, and
proton $p_{T}$ spectra.  
We obtain 
$T_{\mathrm {kin}}$ = 105 $\pm$ 10 (stat.) $\pm$ 16 (sys.) MeV and
$\langle \beta_T \rangle$ = 0.46$c$ $\pm$ 0.01$c$ (stat.) $\pm$ 0.04$c$ (sys.).
The chemical 
freeze-out 
parameters are extracted
from a thermal model fit to the particle ratios at midrapidity. 
We extract 
$T_{\mathrm {ch}}$ = 151 $\pm$ 2 (stat.) $\pm$ 7 (sys.) MeV and 
$\mu_{\mathrm B}$ = 354 $\pm$ 7 (stat.) $\pm$ 30 (sys.) MeV
for 0--10\% central Au+Au collisions at $\sqrt{s_{NN}} =$ 9.2 GeV.

These results from the lowest 
energy collisions studied up to now at RHIC demonstrate the 
capabilities of the STAR detector to pursue
the proposed Beam Energy Scan.
Large and uniform acceptance for all beam energies in a collider set up, excellent
particle identification (augmented by the inclusion of a full barrel Time-Of-Flight~\cite{tof} 
in addition to the large acceptance TPC), 
and higher statistics will offer significant quantitative and qualitative improvement 
over existing data.
The QCD critical point program at RHIC will allow 
us to extensively explore the QCD phase diagram.
It will also allow us to search for 
the onset of various observations related to
partonic matter that have already been uncovered at the highest RHIC energies.

We thank the RHIC Operations Group and RCF at BNL, and the NERSC Center
at LBNL and the resources provided by the Open Science Grid consortium
for their support. This work was supported in part by the Offices of NP
and HEP within the U.S. DOE Office of Science, the U.S. NSF, the Sloan 
Foundation, the DFG cluster of excellence `Origin and Structure of the Universe',
CNRS/IN2P3, RA, RPL, and EMN of France, STFC and EPSRC of the United Kingdom, FAPESP
of Brazil, the Russian Ministry of Sci. and Tech., the NNSFC, CAS, MoST,
and MoE of China, IRP and GA of the Czech Republic, FOM of the 
Netherlands, DAE, DST, and CSIR of the Government of India,
the Polish State Committee for Scientific Research,  and the Korea Sci. \& Eng. Foundation
and Korea Research Foundation.

\end{document}